\documentclass[useAMS]{mn2e}

\usepackage{ulem} 
\usepackage{graphicx}
\usepackage{amsmath}
\usepackage{ulem}
\usepackage{rotating}
\usepackage{amssymb}

\def\lapprox{\hbox{\lower .8ex\hbox{$\,\buildrel < \over\sim\,$}}}
\def\gapprox{\hbox{\lower .8ex\hbox{$\,\buildrel > \over\sim\,$}}}

\title[\textit{HST\/} observations of WISE J104915.57$-$531906.1]{
  \textit{Hubble Space Telescope} astrometry of the closest
  brown dwarf binary system. I. Overview and improved orbit\thanks{
Based on observations with the  NASA/ESA {\it Hubble Space Telescope},
obtained at the  Space Telescope Science Institute,  which is operated
by  AURA, Inc.,  under NASA  contract NAS  5-26555, under  GO-13748 and
GO-14330.
}
}
\author[L.\,R.\,Bedin et al.]{
  L.\,R.\,Bedin$^{1}$\thanks{E-mail: luigi.bedin@oapd.inaf.it}, 
  D.\,Pourbaix$^{2}$\thanks{Senior Research Associate, F.R.S.-FNRS, Belgium.},
  D.\,Apai$^{3,4}$,
  A.\,J.\,Burgasser$^{5}$,
  E.\, Buenzli$^{6}$, \newauthor
  ~H.\,M.\,J.\,Boffin${^7}$, 
  and M.\,Libralato$^{1,8,9}$\thanks{Starting a PostDoc at STScI on July 1st, 2017}
\\  
$^{1}$INAF-Osservatorio Astronomico di Padova, Vicolo dell'Osservatorio 5, I-35122 Padova, Italy\\
$^{2}$Institut d'Astronomie et d'Astrophysique, Universit\'e Libre de Bruxelles (ULB), 1050 Bruxelles, Belgium\\
$^{3}$Department of Astronomy and Steward Observatory, The University of Arizona, 933 N. Cherry Avenue, Tucson, AZ 85721, USA\\
$^{4}$Lunar and Planetary Laboratory, The University of Arizona, 1640 E. University Blvd., Tucson, AZ 85721, USA\\
$^{5}$Center for Astrophysics and Space Science, University of California San Diego, La Jolla, CA 92093, USA\\
$^{6}$Institute for Astronomy, ETH Zurich, Wolfgang-Pauli-Strasse 27, 8093 Zurich, Switzerland\\
$^{7}$European Southern Observatory, Karl Schwarzschild Strasse 2, D-85748 Garching, Germany\\
$^{8}$Space Telescope Science Institute, 3800 San Martin Drive, Baltimore,  MD 21218, USA\\
$^{9}$Dipartimento do Fisica e Astronomia, Univ.\ di Padova, Vicolo dell'Osservatorio 3, I-35122 Padova, Italy\\
}

\begin{document} 

\date{Accepted 2017 May 9. Received 2017 May 9; in original form 2017 April 13}

\pagerange{\pageref{firstpage}--\pageref{lastpage}} \pubyear{201X}

\maketitle
 
\label{firstpage}

\begin{abstract}
Located     at      2\,pc,     the     L7.5+T0.5      dwarfs     system
WISE\,J104915.57\-$-$531906.1  (Luhman\,16\,AB) is  the third  closest
system  known  to  Earth,  making  it a  key  benchmark  for  detailed
investigation   of  brown   dwarf   atmospheric  properties,   thermal
evolution, multiplicity, and planet-hosting frequency.
In  the  first  study  of   this  series  -- based  on  a  multi-cycle
\textit{Hubble Space Telescope} (\textit{HST\,}) program -- we provide
an  overview  of  the  project   and  present  improved  estimates  of
positions,  proper  motions,  annual  parallax, mass  ratio,  and  the
current best assessment of the orbital parameters of the A-B pair.
Our \textit{HST} observations encompass the apparent periastron of the
binary at 220.5$\pm$0.2\,mas at epoch 2016.402.
Although our  data seem  to be  inconsistent with  recent ground-based
astrometric measurements, we also exclude the presence of third bodies
down to Neptune masses and periods longer than a year.
\end{abstract}

\begin{keywords}
  astrometry -- binaries: visual -- brown dwarfs. 
\end{keywords}

%
\section{Introduction}
\label{introduction}
%

The first completeness-corrected planet occurrence rates emerging from
the \textit{Kepler} mission reveal a  larger number of short-period Earth-sized
planets around M dwarfs than around earlier-type (FGK) stars (Dressing
\&   Charbonneau  2013,   Fressin   et  al.\   2013).  Combined   with
state-of-the-art planet  radius--mass relationships,  studies indicate
that  there  are  about  3.5$\times$  as  many  1-4\,$\mathcal{M}_{\rm
  Earth}$ mass planets  around M-type stars than  around G-type stars,
with  a  period-dependence  on  planet  occurrence  rate  that  varies
monotonically with  host spectral types  (Mulders et al.\  2015).  The
larger number of small  planets around M-dwarfs contains \textit{more}
mass in  solids than  the same small  planet population  around G-type
stars  (Mulders et  al.\ 2016).  Given that  discs around  higher-mass
stars have  higher masses  (e.g., Pascucci et  al.\ 2016),  the higher
occurrence  rates  and  higher  solid mass  of  small  planets  around
lower-mass stars may  be the result of inward migration  of planets or
their  planetary  building  blocks.  The  recent  discovery  of  seven
approximately Earth-sized planets in the  TRAPPIST-1 system (a star at
the stellar/substellar boundary) provides  a striking demonstration of
the high occurrence rates of  small planets around small stars (Gillon
et al.\ 2017).

These results motivate  studies of the small  planet population around
even lower-mass hosts, the brown dwarfs (BDs, Kumar 1962, Hayashi \& Nakano
1963).   These low-mass  objects, incapable  of hydrogen  fusion, also
host  circumstellar discs  (e.g., Luhman  et al.\  2008) which  evolve
through the first steps of planet  formation (e.g., Apai et al.\ 2005,
Pascucci et al.\ 2009, Ricci  et al.\ 2013). However, planet detection
through radial  velocity (RV), transit,  and high-contrast imaging  are not
effective  for BD primaries  due to  their low  luminosities
($<$10$^{-3}$~L$_{\odot}$; Burrows et al.\ 2001).
In   contrast,  high-precision   astrometric  observations   have  the
potential to provide an extremely sensitive assessment of the presence
of planets and  to constrain the orbital inclination  of the planetary
system.

The \textit{Wide-field Infrared Survey Explorer (WISE)} has discovered
many  of the  nearest BDs in  the Solar  neighbourhood
(d$<$20\,pc).
A key example is  WISE J104915.57$-$531906.1\,AB, identified by Luhman
(2013) as  a binary BD located  2 pc from the  Sun (hereafter
Luhman\,16\,AB    to   follow    the    Burgasser    et   al.\    2013
denomination). Luhman\,16\,A and B orbit each other at a distance of a
few Astronomical Units with an orbital period  of decades. As the closest
known BDs, the system is ideally suited for detailed characterization.

The primary  (Luhman\,16\,A) is  of spectral  type L7.5$\pm$1  and the
secondary  (Luhman\,16\,B) of  type  T0.5$\pm$1.5.  Both at  effective
temperatures of  about 1,300 K,  placing them near the  L-T transition
(Luhman 2013, Burgasser et al.\ 2013; Kniazev et al.\ 2013; Faherty et
al.\ 2014; Lodieu et al.\ 2015).
The system  age is  constrained to  about 0.1-3\,Gyr,  implying masses
below  0.06\,$\mathcal{M}_{\odot}$ (Faherty  et al.\  2014, Lodieu  et
al.\ 2015). There is no evidence  for the pair belonging to any nearby
young moving  group (Mamajek  2013, Lodieu et  al.\ 2015).  Both brown
dwarfs  are  known  variables,  with the  B  component  more  strongly
variable than A. The variability  likely originates from patchy clouds
(Gillon et al.\  2013, Crossfield et al.\ 2014, Buenzli  et al.\ 2014,
2015a, Karalidi et al.\ 2016).

Boffin  et  al.\ (2014)  reported  perturbations  of the  A-B  orbital
motions in the  Luhman\,16 system, suggesting the presence  of a third
body. Later, Sahlmann \& Lazorenko  (2015) using the same \textit{Very
  Large  Telescope   (VLT)\,}  data  and  those   from  the  follow-up
monitoring programme by  Boffin et al., have excluded  the presence of
any third object with a mass  greater than two Jupiter masses orbiting
around either BD with a period between 20 and 300\,d.

Past   and  present-day   ground-based  seeing-limited   imaging-  and
adaptive-optics-facilities  have  fundamental  limitations  (field  of
view,  PSF  stability,  differential  atmospheric  chromatic  effects,
seasonal visibility),  all of which introduce  systematic and seasonal
astrometric  errors that  are difficult  to quantify  or isolate  when
constraining the presence of companions via astrometry.
This is  particularly true for faint  and red objects, which  are much
redder than the stars used as reference  in the field.  In the case of
Luhman\,16, there is the additional  complication of observing a tight
binary system  (at an  average separation  of $\sim$1$^{\prime\prime}$
and down to 0.22$^{\prime\prime}$, see next sections).
In the past, these systematic errors have resulted in false detections
of planetary companions.
These include reports of exoplanets orbiting Lalande\,21185 (van de
Kamp \& Lippincott 1951, Lippincott 1960) and Barndard's star (van de
Kamp 1963, 1969), both later refuted (Gatewood \& Eichhorn 1973,
Gatewood 1974).
More recently, a giant planet was claimed to orbit the M8 dwarf VB\,10
by  Pravdo  \&  Shaklan   (2009)  based  on  ground-based  astrometric
observations,   a  claim   subsequently   refuted   by  Lazorenko   et
al.\  (2011), among  others. These  and  other examples  point to  the
importance  of high-precision--sub-milli-arcsecond  (mas)--space-based
astrometry  for robust  detection of  exoplanets around  very low-mass
stars and BDs.

For these  reasons, we have  used the \textit{Hubble  Space Telescope}
(\textit{HST\,}) in a special mode  to obtain the most accurate annual
parallax  of any  BD  to date  (eventually  down  to the  50
micro-arc second  level, $\mu$as)  for each of  the two  components of
Luhman\,16, and to constrain their absolute space motions with similar
accuracy.
Most importantly, by searching for astrometric perturbation of the A-B
orbital  motion, we  will  be  able to  confirm  whether giant  planet
candidates exist  in this system  uncovering exoplanets down to  a few
Earth-masses, as described below.
Our  \textit{HST} data  could also  potentially complement/extend  the
work  done by  Melso et  al.\ (2015)  searching for  giant planets  as
resolved,  faint companions  comoving with  the targets;  however, we
will see that this is not the case.

In this  first article, we focus  on the standard imaging  analysis of
our existing  \textit{HST\,} data using procedures  and methods widely
used    in   literature,    sufficient   to    significantly   improve
characterization of this brown-dwarf binary system.
In  Section\,2  we  detail  our observing  strategy.  In  Sect.\,3  we
describe our data reduction and measurements.
We explore two  methods to derive astrometric  and orbital parameters,
through simultaneous fit of the  parameters (Section\,4) and through a
two-step fitting procedure (Section 5).
In  Section\,6  we use  the  limited  radial  velocities for  the  two
components available in the literature to remove the degeneracy in the
sign of the orbital inclination.
In Sect.\,7 we examine the photometric variability of the sources.
In Sec.\  8 we examine the  potential presence of exoplanets  based on
this analysis,  ruling out the  presence of planets more  massive than
one Neptune-mass,  and pointing out several-mas  level inconsistencies
between  \textit{HST\,}  and  existing  ground-based  astrometry  from
Sahlmann \& Lazorenko (2015, hereafter SL15).
In Sect.\,9, we summarize the  electronic material released as part of
this work.  In Sect.\,10 we summarize our conclusions.


%
\section{Observing Strategy}
%

The  imaging data  acquired for  this  project are  obtained with  the
\textit{Ultraviolet-VISual (UVIS)}  channel of the  \textit{Wide Field
  Camera  3  (WFC3)}  instrument   on  \textit{HST\,}  under  programs
GO-13748 and GO-14330 (PI: Bedin).
UVIS has a wide field  of view of about 160$\times$160\,arcsec$^2$ and
the best image-quality on-board  of \textit{HST\,}, with two detectors
of   $\sim$4,000$\times$2,000\,pixels$^2$,  and   a  pixel   scale  of
$\sim$39.77\,mas.

The  main goal  of  our program  is to  probe  Luhman\,16\,AB for  the
presence    of   additional,    yet    unknown    bodies,   down    to
$\sim$5\,Earth-masses,  through astrometric  perturbations of  the A-B
orbital motion.
Our  observations   were  designed  to  maximize   the  high-precision
astrometric capabilities  of the \textit{HST} with  the relatively new
cutting-edge technique of spatial-scanning mode.
Spatial  scanning  under  Fine  Guide  Sensors  (FGS)  control  is  an
observing  mode implemented  for WFC3  on \textit{HST}  only recently,
with the  original aim  of high-precision photometry  during exoplanet
transits (McCullough \& MacKenty 2012).
In this mode the target field is observed while the telescope slews in
a specified  direction and rate.  This spatial scanning enables  up to
thousands  of times  more sampling  of the  same sources  (targets and
references), boosting photometric precision over pointed imaging.
In the following we will refer to these images in spatial-scanning
mode as \textit{trailed}, and to those obtained in the standard mode
as \textit{pointed}.
Recently,  Riess   et  al.\  (2014)  employed   this  newly  developed
\textit{HST} observing mode  for WFC3 also to the  case of astrometry.
By scanning perpendicularly to the  long axis of the parallax ellipse,
this  mode   can  considerably  improve  the   precision  of  parallax
measurements. Riess  et al.\  demonstrated the possibility  to measure
changes in a  source's position to a precision of  20-40 $\mu$as. This
is  almost  an order  of  magnitude  better  than what  is  attainable
employing the best techniques in  traditional pointed imaging with the
WFC3/UVIS (i.e., 320\,$\mu$as, see Bellini, Anderson \& Bedin, 2011).
As a rule of thumb,  the astrometric precision essentially scales with
the length  of the trails. As  in point-source imaging all  the weight
comes  from  the  innermost  5-pixels, therefore  trailing  for  2,000
pixels, results in a 2000/5 = 400 times more pixels, and therefore the
gain is a factor of $\sqrt{400}=20$.
Accurate calculations are more complicated, and residuals in geometric
distortion,  currently calibrated  down  to  320\,$\mu$as (Bellini  et
al.\ 2011)  need to be suppressed  to make this potential  gain usable
down to 20\,$\mu$as (Casertano et al.\ 2016).
There  are  important  differences  between  our  observing  mode  and
analyses, and those  proposed by Riess et al.\ (2014)  or Casertano et
al.\ (2016), which  will be discussed in a  subsequent paper analysing
the trailed images.\\

The data for  this project are collected in 13  visits at well-defined
epochs,  5 at  the  maximum  elongations of  the  parallax,  and 8  at
fractional phases  of the year at  (approximate) logarithmic sampling.
Sampling periods  are between 30\,days  and 3\,years. Twelve  of these
visits have been collected.

The images (two-dimensional data) collected in thirteen epochs provide
26 independent data points to solve for the Luhman\,16\,AB baricentric
positions, barycenter motions, parallax, mass-ratio, and for the seven
parameters of  the A-B  orbit (13 parameters),  plus detection  of any
significant deviation.
Naturally, \textit{it would  not be possible} to  firmly constrain the
A-B orbit until at  least half of the orbit will  be completed, but we
would still  be able  to firmly  detect deviations  from a  conic arc,
induced by  additional bodies, at  least for periods  between $\sim$30
days up to $\sim$3 years.

The roll angle (hereafter, PA$_{\rm  V3}$, i.e., the position angle of
axis  V3 of  the \textit{HST'}s  focal plane\footnote{WFC3  Instrument
  Handbook Chapter\,2.2, Dressel  et al. 2017.}) was  varied among the
visits to minimize hidden systematic  errors.  Data were acquired in 6
different roll-angle values, with each roll-angle used on at least two
visits.
The dither pattern  within each epoch was also  carefully designed. We
imposed  large dither-steps  of at  least $\pm$100\,pixels  to have  a
check   on  the   distortion   residuals.  We   also  avoided   having
Luhman\,16\,A and B fall on any  of the lithographic features or other
known  cosmetic  defects  of  the   two  UVIS  CCDs  (see  Bellini  et
al.\  2011),  or  in  the  gap   between  the  two  chips.  Two  known
extra-galactic sources also fell in all  images, which will be used in
future works as an external check on absolute positions.

The  main  astrometric  trailed  exposures  are  collected  in  filter
WFC3/UVIS/F814W.  Although a  medium or  narrow band  filter would  be
better for astrometry, as PSFs would be less dependent on the color of
the  targets,  filter  F814W  has  the  valuable  property  that  both
Luhman16\,A and B have almost the same count rates.
Furthermore,  the  F814W filter  has  one  of the  best  characterized
geometric distortion solutions (Bellini, Anderson \& Bedin 2011).

Future  analyses  of the  trailed  images  require pointed  images  to
characterize  the  sources  in  the  patch of  sky  (targets  and  the
reference stars)  and color  information to trace  potential chromatic
dependencies.
Therefore, two 60\,s  short pointed exposures in F814W  are also taken
at the  beginning and  at the end  of each orbit  to provide  an input
list, where  both components of  Luhman\,16 are just  below saturation
level, i.e., at the maximum astrometric S/N possible.
Next, to  get the color  information of the  sources in the  field, we
choose the  filter WFC3/UVIS/F606W,  which is sufficiently  bluer than
F814W, and the best compromise between depth and reasonable signal for
the cool components of Luhman\,16.
Due to the  time required by the frame buffer  dump, the exposure time
for an optimal duty cycle  is $\sim$350\,s. Our trailed exposures have
at  least this  exposure times,  and when  visibility allows  it, even
longer.  Therefore, in  addition to the two short  exposures, we could
not fit  more than five  long exposures per  orbit, of which  four are
trailed in F814W, and one is a pointed image in F606W.
%
Seven exposures  per orbits  means that  in the end  there will  be 52
trailed images of at least 350s in F814W, and 39 pointed images, 26 of
which of  60\,s in F814W  and 13 of $\sim$350s  in F606W, for  a grand
total of 91 images.
So far 12 of 13 visits have been collected, i.e., 36 pointed images
are available.

%
\begin{table}
\caption{\textit{HST} images used in this work (ID is not in MJD order).}
\center
\begin{tabular}{lcc}
\hline
\#ID: MJD$_{\rm start}$ & image~ EXPT & PA$_{\rm V3}[^\circ]$ \\
\hline
 & & \\
 F814W & & \\
 & & \\
01: 56891.03123284 & \texttt{icmw09v1q}~ 60\,s & 345.010712 \\ 
02: 56891.10048062 & \texttt{icmw09waq}~ 60\,s & 345.006592 \\
03: 56935.09556248 & \texttt{icmw04l1q}~ 60\,s &  40.016430 \\
04: 56935.12740285 & \texttt{icmw04lcq}~ 60\,s &  40.013191 \\
05: 57031.75281042 & \texttt{icmw02zlq}~ 60\,s & 144.999603 \\
06: 57031.82073857 & \texttt{icmw02zwq}~ 60\,s & 145.003006 \\
07: 57059.93495262 & \texttt{icmw12asq}~ 60\,s & 160.003403 \\
08: 57059.96679262 & \texttt{icmw12b3q}~ 60\,s & 160.007401 \\
09: 57115.70210654 & \texttt{icmw07ccq}~ 60\,s & 215.013794 \\
10: 57115.73394691 & \texttt{icmw07cnq}~ 60\,s & 215.017197 \\
11: 57204.00903025 & \texttt{icmw01vbq}~ 60\,s & 325.005707 \\
12: 57204.07274544 & \texttt{icmw01vyq}~ 60\,s & 325.002289 \\
13: 57255.02930926 & \texttt{icmw10bsq}~ 60\,s & 345.010712 \\
14: 57255.06114963 & \texttt{icmw10caq}~ 60\,s & 345.006592 \\
15: 57300.55774736 & \texttt{icmw05dfq}~ 60\,s &  40.016430 \\
16: 57300.59198366 & \texttt{icmw05dqq}~ 60\,s &  40.013191 \\
17: 57395.65852572 & \texttt{icmw03xoq}~ 60\,s & 144.999603 \\
18: 57395.69036609 & \texttt{icmw03xzq}~ 60\,s & 145.003006 \\
19: 57429.21396601 & \texttt{icmw13uyq}~ 60\,s & 161.070297 \\
20: 57429.28672083 & \texttt{icmw13v9q}~ 60\,s & 161.074203 \\
21: 57479.56281007 & \texttt{icmw08bvq}~ 60\,s & 215.013794 \\
22: 57479.62306489 & \texttt{icmw08c6q}~ 60\,s & 215.017197 \\
23: 57665.08973572 & \texttt{icte06rvq}~ 60\,s &  40.016430 \\
24: 57665.12157609 & \texttt{icte06saq}~ 60\,s &  40.013191 \\
 & & \\
 F606W  & & \\
 & & \\                                
25: 56891.04534173 & \texttt{icmw09v8q}~ 348\,s & 345.008606 \\
26: 56935.10967137 & \texttt{icmw04l6q}~ 348\,s &  40.014809 \\
27: 57031.76691931 & \texttt{icmw02zqq}~ 348\,s & 145.001297 \\
28: 57059.94906114 & \texttt{icmw12axq}~ 348\,s & 160.005402 \\
29: 57115.71621543 & \texttt{icmw07chq}~ 348\,s & 215.015503 \\
30: 57204.02463210 & \texttt{icmw01vhq}~ 348\,s & 325.003998 \\
31: 57255.04341815 & \texttt{icmw10c2q}~ 348\,s & 345.008606 \\
32: 57300.57305995 & \texttt{icmw05dkq}~ 348\,s &  40.014809 \\
33: 57395.67263424 & \texttt{icmw03xtq}~ 348\,s & 145.001297 \\
34: 57429.25623453 & \texttt{icmw13v3q}~ 348\,s & 161.072296 \\
35: 57479.57691896 & \texttt{icmw08c0q}~ 348\,s & 215.015503 \\
36: 57665.10384461 & \texttt{icte06s3q}~ 348\,s &  40.014809 \\
 & & \\
\hline
\end{tabular}
\label{tabimg}
\end{table} 
%

This article is based ---exclusively--- on these pointed images, 24 in
filter F814W and 12 in filter F606W.
Table\,\ref{tabimg} gives information on these 36 images.
%

\section{Data Reductions and Measurements}

In this section we provide a brief description on how the positions in
pixel coordinates $(x,y)$  for all the stars in  the individual frames
were obtained, transformed into a  common reference frame $(X,Y)$, and
into  a standard  equatorial  coordinate  system $(\alpha,\delta)$  at
Equinox J2000. At each step we also give reference to works containing
more exhaustive descriptions of the adopted procedures and software.

\subsection{Correction for imperfect CTE }

Imperfections  in  the  charge  transfer efficiency  (CTE)  smear  the
images, which result in compromised  astrometry (see Anderson \& Bedin
2010).
In all our observations we have mitigated the CTE effects in two ways:
(1)  \textit{passive   CTE-mitigation:  }   we  have   downloaded  the
\texttt{\_flc}  images, which  apply  the  pixel-based CTE  correction
algorithms developed for the \textit{Wide  Field Channel} (WFC) of the
\textit{Advanced Camera  for Surveys}  (ACS) (Anderson \&  Bedin 2010)
and are  already implemented  also for WFC3/UVIS  images by  the STScI
pipeline. They are  now downloadable as standard  data-products at the
MAST archive.\footnote{
  The Mikulski Archive for Space Telescopes (MAST) is a NASA funded project.
  MAST is located at the Space Telescope Science Institute (STScI).
  \texttt{http://archive.stsci.edu/hst}
}
(2) \textit{active  CTE-mitigation: } We have  applied a post-flashing
(of  about  12  e$^-$)  to  keep the  background  above  the  critical
threshold (filling many  of the charge-traps), which  suppress as much
as     convenient      the     residuals     due      to     imperfect
CTE.\footnote{\texttt{http://www.stsci.edu/hst/wfc3/ins\_performance/CTE/}}
However, both  strategies do not  work perfectly, and traces  of these
imperfect CTE remain. The residuals left on the measured positions are
sizable ($\sim$0.5\,mas),  but thankfully, they are  also (relatively)
easy to track down and remove (see following Sections).

\subsection{Fluxes and positions in the individual images}

Positions and fluxes of sources in each WFC3/UVIS \texttt{\_flc} image
were  obtained  with  software  that   is  adapted  from  the  program
\texttt{img2xym\_WFC.09x10}  developed for  ACS/WFC (Anderson  \& King
2006), and publicly available.\footnote{
\texttt{http://www.stsci.edu/$\sim$jayander/WFC3/}
}
Together with the  software, a library of  \textit{effective} PSFs for
most common filters is also  released. These are spatially variable in
a 7$\times$8 array, and can also  be perturbed in a spatially constant
mode to better fit PSFs of individual frames.
However, we follow the Bellini  et al.\ (2013) prescription to perturb
the  library PSFs  also  spatially (in  a  5$\times$5 spatial  array).
These procedures tailor the library PSFs to each individual image even
better than spatially-constant perturbed  PSFs, as they better account
for small focus variations across the whole field of view.
In addition  to solving  for positions and  fluxes, the  software also
provides   a  quality-of-fit   parameter  ($Q$).   The  quality-of-fit
essentially tells how  well the flux distribution  resembles the shape
of the point spread function (this parameter is defined in Anderson et
al.\  2008).  It  is  close  to zero  for  stars  measured best.  This
parameter  is  useful  for  eliminating galaxies,  blends,  and  stars
compromised by detector cosmetic or artifacts.

Once the raw pixel positions $(x^{\rm raw},y^{\rm raw})$ and magnitude
are obtained, they are corrected for geometric distortion. We used the
best available  average distortion corrections for  WFC3/UVIS (Bellini
\& Bedin  2009; Bellini, Anderson  \& Bedin  2011) to correct  the raw
positions  and fluxes  of sources  that  we had  measured within  each
individual image.
(Note  that  fluxes  are  also  corrected for  pixel  area  using  the
geometric distortion correction).
We  refer to  corrected positions  in  the individual  frame with  the
symbols $(x^{\rm cor},y^{\rm cor})$.

Finally, we note that given the large width of the filter pass-band of
F606W  and  F814W, the  PSFs  for  very  red stars  are  significantly
different from the PSFs of average-color stars in the field. This fact
is evident in the relatively large values of $Q$ for Luhman\,16\,A and
B,  compared to  stars of  similar  magnitudes; and  residuals in  the
subtracted images of  Luhman\,16\,A and B. These  fit mismatches could
potentially lead to chromatic systematic errors in positions, known to
affect UV filters (Bellini et al.\ 2011). However, as described below,
such offsets  are below  the level of  other systematic  errors (i.e.,
$\sim$320\,$\mu$as for geometric distortion) for this analysis.

%
\begin{table*}
  \caption{Observed positions for Luhman\,16 A and B in both the
    master frame coordinate system and in the raw coordinate
    system of the individual images. The ID is as in Table\,1.}
  \center
\begin{tabular}{ccccccccc}
\hline
\#ID & $X_{\rm A}^{\rm obs}$ & $Y_{\rm A}^{\rm obs}$ & $x_{\rm A}^{\rm raw}$  & $y_{\rm A}^{\rm raw}$ & $X_{\rm B}^{\rm obs}$ & $Y_{\rm B}^{\rm obs}$ & $x_{\rm B}^{\rm raw}$  & $y_{\rm B}^{\rm raw}$ \\
\hline
& & & & & & & & \\
01 & 3177.4275 & 4480.9977  & 2180.594 & 3248.659 & 3200.5454 & 4483.9296 & 2203.846 & 3249.985\\ 
02 & 3177.4303 & 4481.0021  & 2382.559 & 3031.726 & 3200.5414 & 4483.9315 & 2405.762 & 3033.062\\ 
03 & 3176.9366 & 4479.5016  & 2884.488 & 2699.714 & 3198.7634 & 4481.8825 & 2898.892 & 2682.114\\ 
04 & 3176.9435 & 4479.5022  & 3084.968 & 2484.191 & 3198.7643 & 4481.8713 & 3099.326 & 2466.629\\ 
05 & 3184.4565 & 4459.9769  & 2196.945 & 1018.424 & 3203.3583 & 4461.0975 & 2179.580 & 1012.017\\ 
06 & 3184.4761 & 4459.9582  & 2397.461 &  805.277 & 3203.3756 & 4461.0672 & 2380.121 &  798.883\\ 
07 & 3192.5751 & 4453.7701  & 1675.234 & 1020.767 & 3210.5832 & 4454.5311 & 1657.325 & 1019.570\\ 
08 & 3192.5676 & 4453.7692  & 1876.887 &  807.325 & 3210.5920 & 4454.5209 & 1858.992 &  806.127\\ 
09 & 3214.1073 & 4447.1471  &  978.149 & 1568.796 & 3230.3335 & 4447.1769 &  967.574 & 1581.945\\ 
10 & 3214.1009 & 4447.1548  & 1181.296 & 1353.836 & 3230.3353 & 4447.1909 & 1170.739 & 1366.957\\ 
11 & 3242.2848 & 4452.4870  & 1717.135 & 3241.691 & 3255.6215 & 4451.3658 & 1730.160 & 3244.327\\ 
12 & 3242.3069 & 4452.4778  & 1920.016 & 3024.247 & 3255.6413 & 4451.3684 & 1933.013 & 3026.896\\ 
13 & 3247.3853 & 4456.4204  & 2247.516 & 3213.591 & 3258.9968 & 4454.6607 & 2259.184 & 3210.991\\ 
14 & 3247.3841 & 4456.4230  & 2449.420 & 2996.722 & 3259.0010 & 4454.6512 & 2461.070 & 2994.122\\ 
15 & 3247.0263 & 4454.9574  & 2900.188 & 2632.622 & 3257.0703 & 4452.6038 & 2903.983 & 2622.733\\ 
16 & 3247.0203 & 4454.9533  & 3100.615 & 2417.148 & 3257.0684 & 4452.5891 & 3104.392 & 2407.270\\ 
17 & 3254.4639 & 4435.7435  & 2131.405 & 1031.633 & 3261.1672 & 4432.1937 & 2123.896 & 1033.147\\ 
18 & 3254.4433 & 4435.7412  & 2332.359 &  818.548 & 3261.1601 & 4432.1852 & 2324.849 &  820.057\\ 
19 & 3264.2778 & 4428.4527  & 1594.860 & 1046.116 & 3269.8193 & 4424.4668 & 1589.037 & 1050.078\\ 
20 & 3264.3006 & 4428.4303  & 1796.647 &  832.784 & 3269.8390 & 4424.4459 & 1790.841 &  836.733\\ 
21 & 3283.9823 & 4422.6438  &  939.376 & 1641.428 & 3287.7122 & 4418.0588 &  940.481 & 1647.189\\ 
22 & 3283.9962 & 4422.6308  & 1142.653 & 1426.469 & 3287.7264 & 4418.0350 & 1143.768 & 1432.227\\ 
23 & 3317.6832 & 4430.3163  & 2920.286 & 2558.888 & 3314.7398 & 4423.5832 & 2913.082 & 2557.926\\ 
24 & 3317.6947 & 4430.3326  & 3120.572 & 2343.566 & 3314.7544 & 4423.5828 & 3113.368 & 2342.590\\ 
& & & & & & & & \\                                                                                               
25 & 3177.4460 & 4481.0101  & 2281.575 & 3139.634 & 3200.5467 & 4483.9358 & 2304.789 & 3140.960\\ 
26 & 3176.9415 & 4479.5348  & 2984.632 & 2591.444 & 3198.7467 & 4481.9008 & 2998.996 & 2573.871\\ 
27 & 3184.4592 & 4459.9705  & 2297.332 &  911.318 & 3203.3471 & 4461.0937 & 2279.990 &  904.925\\ 
28 & 3192.5569 & 4453.7925  & 1776.067 &  913.452 & 3210.5696 & 4454.5195 & 1758.167 &  912.284\\ 
29 & 3214.1096 & 4447.1664  & 1079.659 & 1460.824 & 3230.3390 & 4447.1832 & 1069.104 & 1473.969\\ 
30 & 3242.3159 & 4452.4941  & 1818.561 & 3132.291 & 3255.6412 & 4451.3722 & 1831.562 & 3134.924\\ 
31 & 3247.3933 & 4456.4421  & 2348.386 & 3104.735 & 3259.0001 & 4454.6511 & 2360.038 & 3102.110\\ 
32 & 3247.0193 & 4454.9672  & 3000.366 & 2524.365 & 3257.0534 & 4452.6149 & 3004.151 & 2514.496\\ 
33 & 3254.4318 & 4435.7819  & 2231.914 &  924.491 & 3261.1580 & 4432.2029 & 2224.381 &  926.023\\ 
34 & 3264.3046 & 4428.4673  & 1695.718 &  938.990 & 3269.8228 & 4424.4705 & 1689.925 &  942.958\\ 
35 & 3283.9613 & 4422.6474  & 1041.177 & 1533.354 & 3287.6950 & 4418.0489 & 1042.291 & 1539.121\\ 
36 & 3317.6784 & 4430.3462  & 3020.289 & 2450.742 & 3314.7539 & 4423.6158 & 3013.104 & 2449.764\\ 
& & & & & & & & \\
\hline
\end{tabular}
\label{tabABXYxy}
\end{table*} 
%

\subsection{The reference frame}

As our  \textit{reference frame}  we adopted the  best-fit ($\sim$200)
stars measured in the first exposure of our first epoch.
These stars were unsaturated, with  at least 4,000\,$e^-$, isolated by
at least  9 pixels, and  with $Q<0.4$.  To avoid negative  values when
stacking  all images  from all  epochs  (see next  section), we  added
1000.0 pixels in each coordinate to the distortion-corrected positions
of these stars;  the resulting coordinates are  indicated with $(X,Y)$
and are the positions in our adopted reference frame.
We then  find the  most general linear  transformation (6-parameters),
between  the  distortion-corrected positions  of  these  stars of  the
reference-frame  $(X,Y)$,  and  their  distortion-corrected  positions
$(x^{\rm cor},y^{\rm cor})$  in any other image (for as  many stars in
common as possible).
This  enables us  to perform  all  of the  relative measurements  with
respect to this reference frame.
These transformations  were computed  using only stars  with positions
consistent to at least 3.6\,mas (or 0.09 pixels). This ensured that we
use only stars that are present in  all imaging epochs and do not have
compromised measurements due to cosmic rays or detector cosmetics.
The underlying assumption  is that stars in the  field have negligible
common and  peculiar motions;  however an  uncertainty of  3.6\,mas on
$\sim$200 stars still  leads to an uncertainty of  250\,$\mu$as on the
centroid position.
In this work we will not  attempt to iteratively solve for the motions
of individual  reference field objects, as  again, these uncertainties
are of  the same order of  the precision of the  astrometry in pointed
images.
We  will see  in Sect.\,\ref{gaia}  how this  positional precision  is
consistent with \textit{Gaia} DR1 positions.

For  the same  reasons, and  differently from  the more  sophisticated
local-transformations procedure described in  Bedin et al.\ (2014), we
do not  apply any local  approach, due  to the relatively  low stellar
density  in  the  field,  and  as our  final  residuals  using  global
transformations   are    already   consistent   with    our   expected
uncertainties.

The   observed  positions   in   the  $(X,Y)$   reference  frame   for
Luhman\,16\,A and B in the 36 pointed images analyzed in this work are
given  in Table\,\ref{tabABXYxy}.  There,  we also  give the  $(x^{\rm
  raw},y^{\rm raw})$  positions which will  be used to track  down the
systematic effects of residuals of  imperfect CTEs, and to correct for
them.

\subsection{Absolute Astrometric Calibration}
\label{gaia}

We  anchor  our reference  frame  to  the  astrometric system  of  the
\textit{Gaia\,}   first    data-release   (DR1,    The   \textit{Gaia}
collaboration, 2016), which  is tied to the ICRS  for equinox J2000.0,
and at epoch 2015.0.

As  extensively  discussed  in  Bedin  et  al.\  (2014),  any  adopted
geometric distortion correction for  \textit{HST\,} cameras is just an
average solution,  as from frame  to frame there are  sizable changes,
mainly  induced by  velocity aberration  (Cox \&  Gilliland 2003)  and
focus variations  (the so  called \textit{breathing} of  the telescope
tube as a result of different incidence of light from the Sun).
This  is particularly  true for  the linear  terms of  the distortion,
which contain the largest portion of these changes.
We  used   six-parameter  linear   transformations  to   register  the
distortion-corrected  positions   measured  in  each   frame  $(x^{\rm
  cor},y^{\rm  cor})$ to  the  distortion-corrected  positions of  the
reference frame  $(X,Y)$. The  linear-term variations with  respect to
the  reference   frame  are   almost  completely  absorbed   by  these
six-parameter linear transformations. However, the linear terms of the
reference  frame remain  to  be  determined: astrometric  zero-points,
plate scales, orientations, and skew terms need to be calibrated to an
absolute  reference system.  To  avoid under-sampling  in our  stacked
image we derive the transformation from \textit{Gaia} to our reference
frame  super-sampled  by  a  factor   of  2,  hereafter  indicated  as
$(2X,2Y)$, see next section.

Within  our  Luhman\,16\,AB  WFC3/UVIS  field of  view  we  found  126
\textit{Gaia\,} DR1 point  sources in common with  our reference frame
which  we used  to calibrate  our six  linear terms.   To do  this, we
assume  our $(2X,2Y)$  as  a  tangential plane  in  the tangent  point
$(\alpha_0,\delta_0)$,            and           matched            the
$(\alpha-\alpha_0,\delta-\delta_0)$ differences with our $(2X,2Y)$. We
then      find       the      linear       transformation      between
$(\alpha-\alpha_0,\delta-\delta_0)$ and $(2X,2Y)$,  by solving for the
six parameters $\mathcal{A,B,C,D},X_0,Y_0$ in the system:
\begin{equation*}
\begin{cases}
\alpha^* = (\mathcal{A}(2X-X_0)+\mathcal{B}(2Y-Y_0))+\alpha_0^* \\ 
\delta = (\mathcal{C}(2X-X_0)+\mathcal{D}(2Y-Y_0))             +\delta_0 \\ 
\end{cases}
\end{equation*}
Where $\alpha^*=\alpha \cos{\delta}$. 
All 126  stars in common  with Gaia DR1  agree to within  40\,mas with
positions in $(2X,2Y)$.  Of the 67 within  1 mas, we noticed a bulk of
very consistent stars. To derive the six coefficients we used only the
21  stars  with consistent  Gaia  positions  to within  0.6\,mas.  The
coefficients of the transformations are given in Table~\ref{tab2X2Y}.

%
\begin{table}
\caption{Adopted coefficients to transform the observed tangential
  plane coordinates of the reference frame $(2X,2Y)$ into equatorial coordinates 
  of Gaia DR1 equinox J2000.0, at epoch 2015.0 $(\alpha,\delta)$, which are linked to the ICRS.
}
\center
\begin{tabular}{lc}
\hline
%
%
$\mathcal{A}$      & $   (-4.79474 \pm 0.00025)E-06 $  \\
$\mathcal{B}$      & $   (+2.74144 \pm 0.00025)E-06 $  \\
$\mathcal{C}$      & $   (+2.74145 \pm 0.00025)E-06 $  \\
$\mathcal{D}$      & $   (+4.79468 \pm 0.00025)E-06 $  \\
$X_0$      & $+6062.888 \pm 0.004$ \\
$Y_0$      & $+6832.005 \pm 0.004$ \\
$\alpha_0$ & $   162.302050 $ (defined)       \\ 
$\delta_0$ & $   -53.328911 $ (defined)       \\
\hline
\end{tabular}
\label{tab2X2Y}
\end{table} 
%

%
The plate-scale  derived from  \textit{Gaia} DR1 of  our super-sampled
reference               frame               $(2X,2Y)$               is
$\sqrt{|{\mathcal{A}\mathcal{D}-\mathcal{B}\mathcal{C}}|}=19.883$\,mas,
meaning that the  assumed pixel scale of $(X,Y)$  is 39.766\,mas. This
is  in  good  agreement  with  values of  the  WFC3/UVIS  pixel  scale
independently derived by Bellini et al.\ (2011).

\begin{figure*}
\begin{center}
\includegraphics[width=88mm]{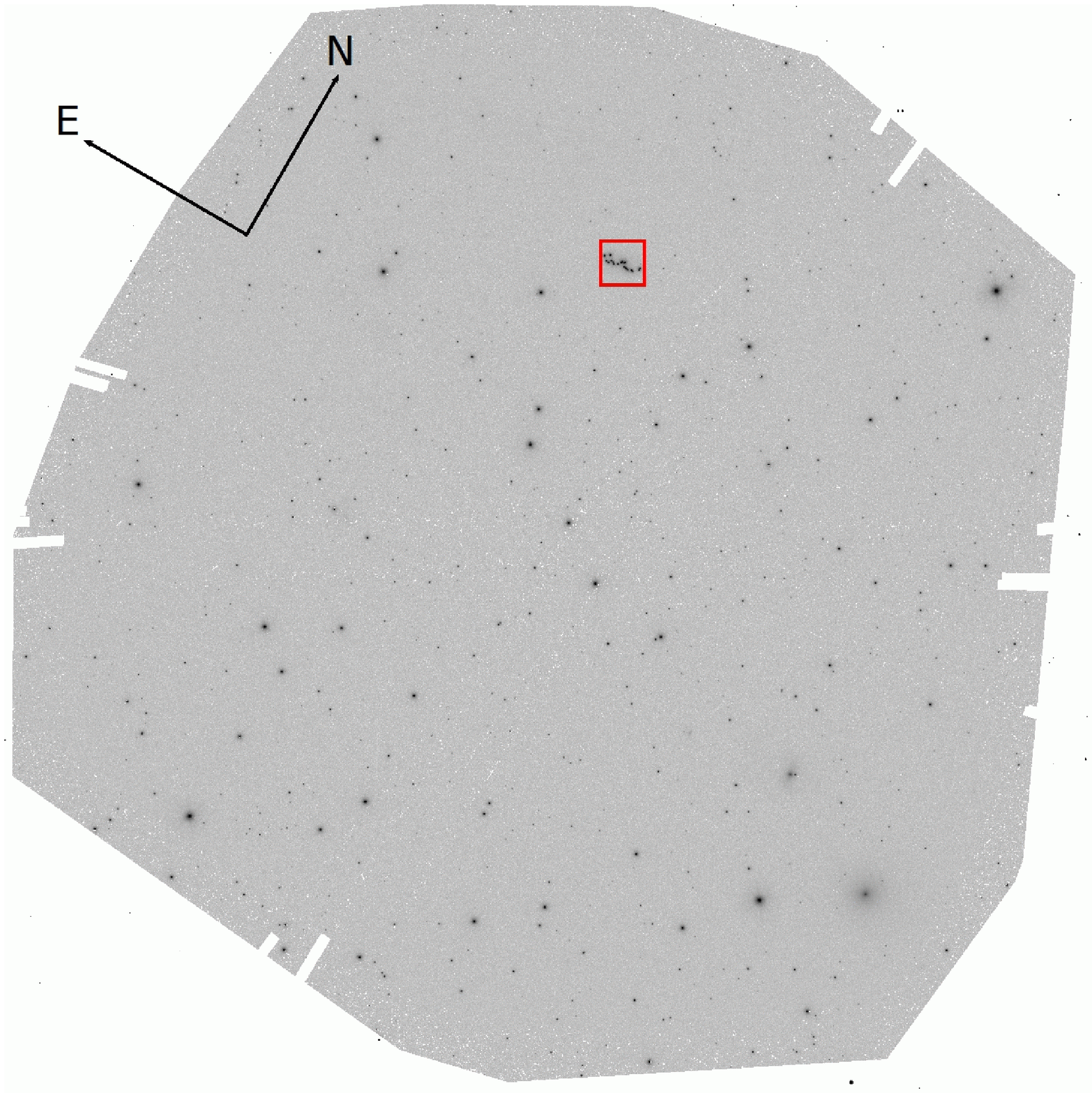}
\includegraphics[width=88mm]{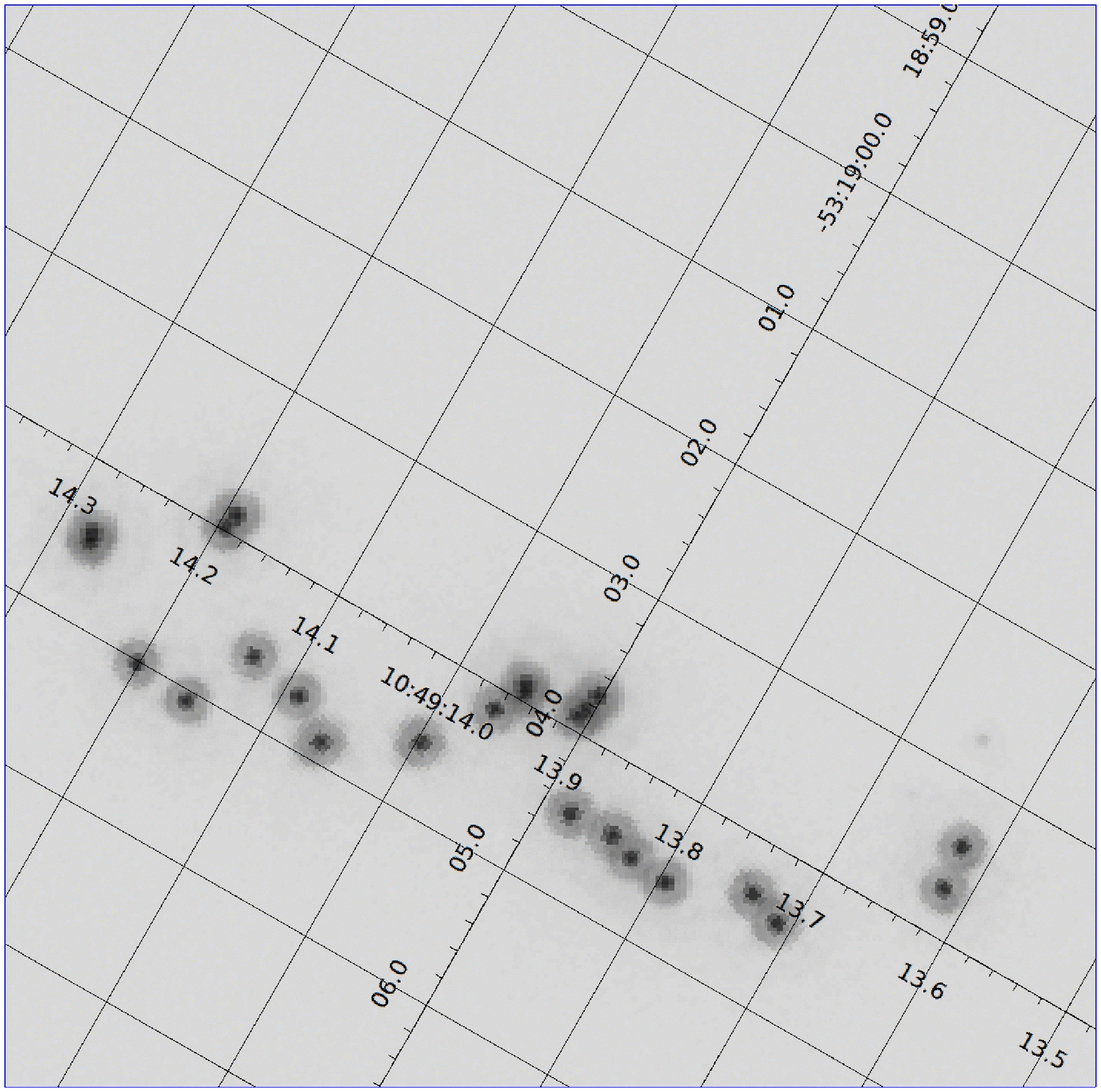} 
\caption{
  \textit{(Left:)} The region surrounding Luhman\,16\,AB, as monitored
  by     \textit{HST\,}.     It     has    dimensions     of     about
  160$^{\prime\prime}$$\times$160$^{\prime\prime}$ and  is the  sum of
  the stacks in WFC3/UVIS/F814W obtained for each of the 12 individual
  visits considered in this work.
   \textit{(Right:)}  Zoom-in  of  the   same  image  in  the  portion
   highlighted   by    a   red   square.     It   has   a    size   of
   $7^{\prime\prime}\times7^{\prime\prime}$  and  shows  the  complete
   pattern  in  the sky  of  Luhman\,16\,A  and  B during  the  period
   monitored by  our \textit{HST\,}  observations. The  orientation is
   the one of the master frame, while a fine grid ($1^{\prime\prime}$)
   gives equatorial coordinates.
\label{stack}
}
\end{center}
\end{figure*}

The non-linear  part of the  WFC3/UVIS distortion solutions  should be
accurate  to  much  less   than  0.01  original-size  WFC3/UVIS  pixel
($\sim$0.3\,mas in a global sense,  Bellini et al.\ 2011), roughly the
random positioning accuracy with which we can measure a bright star in
a single exposure.
Note that  the linear  terms of the  ACS/WFC distortion  solution have
been changing slowly over time (Anderson et al.\ 2007; Ubeda, Kozurina
\& Bedin 2013). Even if this  is the case for WFC3/UVIS, having linked
our  linear terms  to the  \textit{Gaia\,} DR1  makes our  astrometric
solution immune  to those  effects, as  well as  to those  of velocity
aberration, breathing, etc.
Therefore  our   absolute  coordinates   are  referred  to   the  ICRS
\textit{Gaia}  DR1 in  equinox J2000.0,  with positions  given at  the
reference epoch, 2015.0 (The \textit{Gaia} collaboration, 2016).\\

\subsection{Image stack}

Having at  hand all the  transformations from the coordinates  of each
image  into the  reference  frame,  it becomes  possible  to create  a
stacked  image  of the  field  for  each epoch,  and  a  sum of  these
stacks. The stack provides a  representation of the astronomical scene
that enables us  to independently check the region  around each source
at   each   epoch.   The   stacked   images   are   13\,000   $\times$
13\,000\,pixels$^2$   in  the   $(2X,2Y)$   reference  system,   i.e.,
super-sampled  by a  factor of  two ($\sim$20\,mas  pixel$^{-1}$). The
image   sum  of   the  stacks   for  the   12  epochs   is  shown   in
Fig.\,\ref{stack}. We have included in the header of the image (in the
form  of World  Coordinate System  keywords) our  absolute astrometric
solution, which is  based on the \textit{Gaia} DR1  source catalog, as
described in the previous section.  As part of the electronic material
provided in this paper, we also release the stacked average image, and
the sum of  the stacked images for  all epochs in F814W;  all with our
astrometric solution  in the  header in the  form of  World Coordinate
System (WCS).\\

In the  next two sections  we will  describe the determination  of the
astrometric and orbital parameters of the Luh\,16\,AB system employing
two different methods.


\section{METHOD\,A: Simultaneous Determination of Astrometric and Orbital Parameters}

Our reference system  is not an absolute reference frame,  as even the
stars that moved the least and  with the most robust positions are not
at an infinite distance.
On the basis of a Besan\c{c}on  Galactic model (Robin et al.\ 2004), we
expect  that  our best  measured  stars  used  to define  the  $(X,Y)$
reference frame lie at an average distance of $\sim$5\,kpc. This would
introduce a correction from relative  to absolute of about 0.2\,mas to
our parallax and motions.

As noted  previously, in this  first analysis  we aim for  an accuracy
equivalent to  geometric distortion,  about 0.3\,mas. Hence,  to first
approximation  we assume  our measured  positions are  on an  absolute
system.
Rather  than fitting  the relative  orbit (seven  parameters) and  the
absolute motion of the baricenter plus the mass ratio (six parameters)
separately, a  thirteen-parameter model of the  independent motions of
Luhman\,16\,A  and  B  is  adopted.    In  terms  of  complexity,  the
computational  cost  of fitting  both  models  is the  same;  however,
fitting a  model looking  at the separate  component motions  makes it
possible to measure  the correlation between any pair  of the thirteen
parameters, and thus  deviation from that model for  a planet orbiting
one of the components.

The  positions  ($\alpha^*=\alpha\cos\delta$,  $\delta$)  of  the  two
components ($k$) can be expressed as
\begin{equation}\label{eq:orbmodel2}
\left\{
\begin{array}{r@{\quad=\quad}l}
\alpha^*_1 & \alpha^*_0+\varpi f_a+\mu_{\alpha^*}\Delta t+B\,X(t)+G\,Y(t)\\
\delta_1& \delta_0+\varpi f_d+\mu_{\delta}\Delta t+A\,X(t)+F\,Y(t)\\
\alpha^*_2 & \alpha^*_0+\varpi f_a+\mu_{\alpha^*}\Delta t-B\rho X(t)-G\rho Y(t)\\
\delta_2& \delta_0+\varpi f_d+\mu_{\delta}\Delta t-A\rho X(t)-F\rho Y(t)
\end{array}
\right.
\end{equation}
where ($\alpha^*_0$, $\delta_0$) are the position of the baricenter of
the  system at  the reference  epoch (here  2015.0).  $f_a$  and $f_d$
denote  the parallactic  factors  (van\,de\,Kamp  1967), $\varpi$  the
parallax, and $\mu$ the proper  motions along $\alpha^*$ and $\delta$.
For the sake of linearity, we  adopt the Thiele-Innes formalism of the
orbital contribution:
\begin{eqnarray}
X(t)&=&\cos E-e\\
Y(t)&=&\sqrt{1-e^2}\sin E\\
E&=&\frac{2\pi}{P}(t-T_0)+e\sin E\\
A&=&a_1(\cos\omega\cos\Omega-\sin\omega\sin\Omega\cos i)\\
B&=&a_1(\cos\omega\sin\Omega+\sin\omega\cos\Omega\cos i)\\
F&=&a_1(-\sin\omega\cos\Omega-\cos\omega\sin\Omega\cos i)\\
G&=&a_1(-\sin\omega\sin\Omega+\cos\omega\cos\Omega\cos i)
\end{eqnarray}
where $a_1$  is the angular semi-major  axis of the absolute  orbit of
the primary component,  $\omega$ is the argument of  the periastron of
the  primary, $\Omega$  is  the  longitude of  the  node,  $i$ is  the
inclination of the orbital plane  with respect to the plane orthogonal
to the line  of sight, $e$ is  the eccentricity of the  orbit, $P$ the
orbital period and $T_0$ one epoch of periastron passage.

Assuming  that model,  the parameters  which minimize  the sum  of the
residuals  squared  (for  both  components,  both  in  $\alpha^*$  and
$\delta$) are  adopted as the  solution.  Owing  to the nature  of the
model (motion  of the  baricenter and orbit  of both  components), the
gradient of that  sum is nonlinear and one therefore  needs a reliable
initial guess  of the solution  before calling any  local minimization
method (e.g., Levenberg-Marquardt, Marquardt 1963).  Even if our model
is globally nonlinear, the minimization  problem can be reorganized as
two  nested ones,  the inner  one containing  all the  parameters that
appear   linearly  in   the  gradient   of  the   original  sum.    If
$F(p_1,\dots,p_{13})$ denotes that sum  with $p_k$ ($k=1\dots 13$) the
different parameters, one can rewrite the minimization of $F$ as
\begin{eqnarray}\label{equ:pationchi2}
  \min_{(p_1,\dots,p_{13})} F(p_1,\dots,p_{13})= \\
= \min_{(p_{10},\dots,p_{13})}\min_{(p_1,\dots,p_{9})}F(p_1,\dots,p_{9},\underline{p_{10}},\underline{p_{11}},\underline{p_{12}},\underline{p_{13}}) 
\end{eqnarray}
where, for  the sake of simplicity,  it is assumed that  the last four
parameters are those that appear non-linearly in the model (namely the
eccentricity, the orbital period, an epoch of periastron time, and the
scaling factor  $\rho$).  The notation $\underline{p_{k}}$  means that
$p_{k}$ remains unchanged.

In these nested minimizations, the inner  one is linear and thus takes
no iteration  to reach  the minimum.   Therefore, one  ends up  with a
simpler nonlinear model  with four parameters only.   An initial guess
of  that 4-dimensional  nonlinear problem  can be  obtained through  a
basic grid  search before  the whole 13-parameter  nonlinear objective
function is minimized with, say, Levenberg-Marquardt.

The residuals  of this model  exhibit a seasonal variation  along both
axes ($\alpha^*$  and $\delta$) for both  components.  The periodicity
(one  year) and  the  fact that  this variation  is  present in  every
residual rule out the presence of  a companion as an explanation.  If,
instead  of  these  natural  residuals  versus  time,  one  plots  the
residuals in $X$  and $Y$ versus $x_{\rm raw}$ and  $y_{\rm raw}$, the
seasonal  variation  becomes  a  straight  line,  thus  revealing  the
presence of some CTE residual
in the observations.  Four straight  lines are adjusted to correct for
CTE: one for each axis and each filter.
[Note that  at each  of the 13  epochs we have  3 observations  in two
  coordinates, so 72 independent data points, for twelve epochs.]
Once the CTE  residuals are corrected for, a new  model fit is carried
out.   The thirteen  parameters resulting  from that  minimization are
listed in Table~\ref{tab:FullSolution}.  The  typical residual is 0.39
mas.\\

\begin{table}
  \caption[]{\label{tab:FullSolution}    Astrometric    and    orbital
    parameters  of Luhman\,16\,AB  obtained  with  a simultaneous  fit
    described  in Sect.\,4.   The  coordinates are  in J2000.0,  epoch
    2015.0.}
  \begin{tabular}{lcc}\hline
$\alpha^*$ (deg.)$^a$ & +96.9342078 & $\pm$ 7.68E$-$07 \\
$\delta$ (deg.)  & $-$53.3179180 & $\pm$ 5.06E$-$07\\
$\varpi$ (mas)   &  501.14 & $\pm$0.052\\
$\mu_{\alpha^*}$ (mas/yr)   & $-$2763 & $\pm$1.45\\
$\mu_{\delta}$ (mas/yr)    &  +358 & $\pm$1.75\\
$A$ (mas) &   +530 & $\pm$58.8\\
$B$ (mas) & $-$410 & $\pm$348\\
$F$ (mas) & +49 & $\pm$1490\\
$G$ (mas) & $-$183 & $\pm$1250\\
$e$       & 0.25 & $\pm$0.0648\\
$P$ (yr)  & 19 & $\pm$20.2\\
$T_0$ (Jul.yr) & 2000 & $\pm$24.6\\
$\rho$ & +1.22 & $\pm$0.0214\\
    \hline
$^a$~$\alpha^*$ = $\alpha\cos\delta$. &&
  \end{tabular}
\end{table}

Despite the  exquisite precision of  the \textit{HST} data,  there are
difficulties  in   obtaining  a  reliable  simultaneous   fit  of  the
astrometric and  orbital parameters  using only a  partial arc  of the
orbit. It is particularly unsatisfactory  that the orbital period that
results is largely  unconstrained (see Table\,\ref{tab:FullSolution}),
and so too the individual masses.
Similar  difficulties  to  obtain  a simultaneous  fit  were  reported
previously by  Sahlman \&  Lazorenko (2015,  Sect.\,3.1), who  used an
even shorter  orbital arc than  ours.  Among other  inconsistencies in
their  simultaneous solutions,  they noted  a strong  anti-correlation
between the  residuals for  the A  and B  components much  larger than
their measurement uncertainties, indicating that the motion of A and B
were not independent and should be modelled globally.
We therefore analyzed our own A and B motion residuals, and found that
the  Pearson  correlation coefficients  are  $\sim$0.65  for both  the
$\alpha*$ and $\delta$ components of our solution, indeed indicating a
degree of correlation much larger than our expected uncertainties.


\begin{figure*}
\begin{center}
\includegraphics[width=150mm]{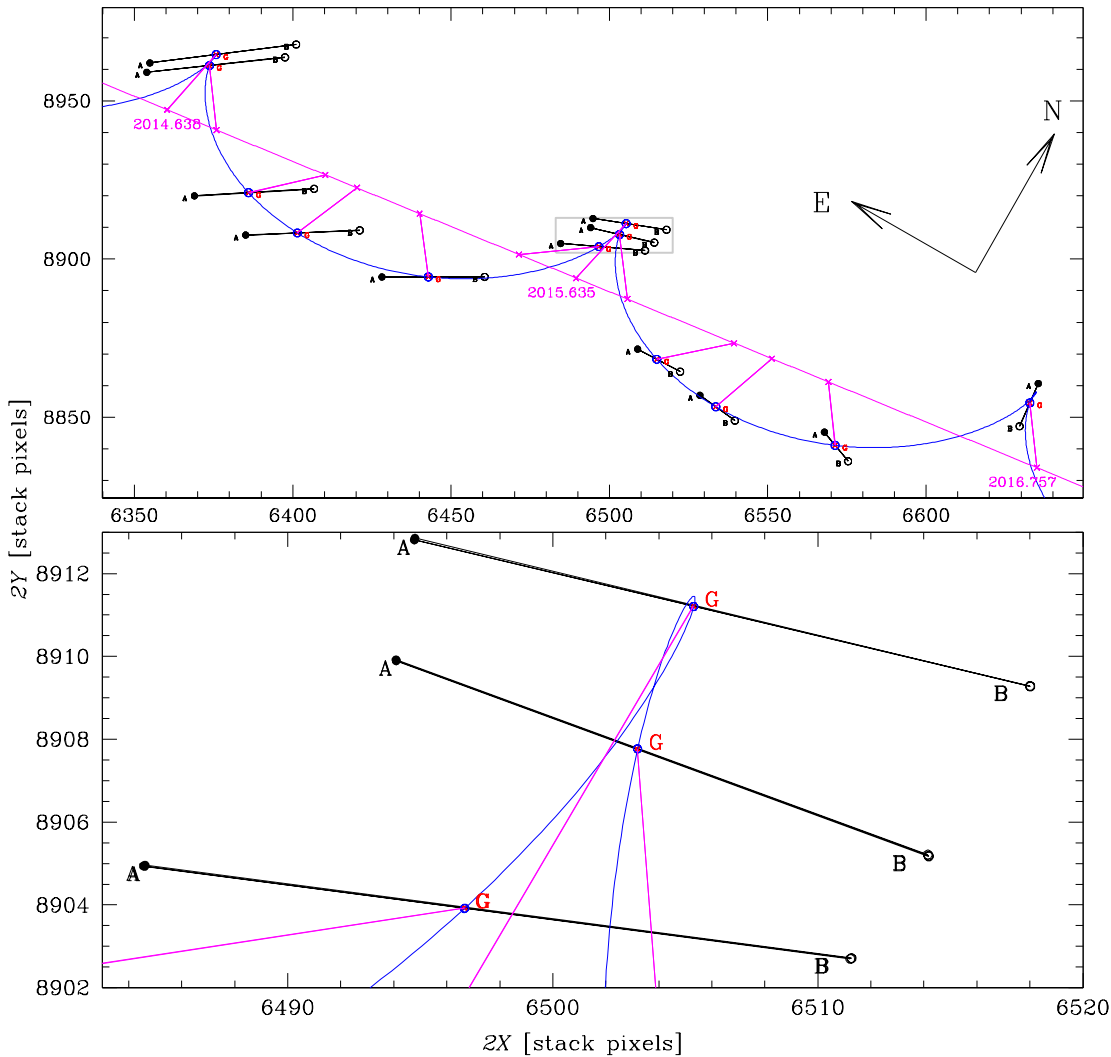}
\caption{
The  top panel  shows the  complete series  of observed  positions for
Luhman\,16\,A (filled  circles) and  Luhman\,16\,B (empty  circles) in
the coordinate system of our master  frame. At any given epoch a black
line  join A  and B,  and along  it  is marked  with a  red cross  the
position of the  baricenter, G, estimated as explained in  the text. A
blue curve indicate the  astrometric solution for the Luhman\,16\,AB's
baricenter (positions, proper motions, parallax), noting that the mass
ratio, $q$, (i.e., the baricenter itself)  is part of the solution.  A
line  in magenta  indicate the  same solution  with the  system at  an
infinite distance,  and crosses in  magenta on this line  indicate the
individual  epochs on  this. These  locations are  connected with  the
observed baricentric positions. The orientations  in this plot are the
same of  the master frame,  as in Fig.\,\ref{stack}. The  bottom panel
shows  details for  these  curves  around epoch  2015.6,  in a  region
indicated in the top panel with a gray rectangle; this panel shows how
finely observations were reproduced.
\label{cycloid}
}
\end{center}
\end{figure*}

%
\begin{table*}
\caption{De-trended coordinates for Luhman\,16 A and B,
  with magnitudes and quality fit parameter.}  \center
\begin{tabular}{ccccccccc}
\hline
\#ID & $X_{\rm A}^{\rm dtr}$ & $Y_{\rm A}^{\rm dtr}$ & $Q_{\rm A}$ & mag.$_{\rm A}$ & $X_{\rm B}^{\rm dtr}$  & $Y_{\rm B}^{\rm dtr}$ & $Q_{\rm B}$ & mag.$_{\rm B}$ \\
\hline
& & & & & & & & \\
01 & 3177.4290 & 4480.9726 & .0595 & $-$13.7789 & 3200.5478 & 4483.9043 & .0678 & $-$13.7686 \\
02 & 3177.4385 & 4480.9813 & .0554 & $-$13.7958 & 3200.5502 & 4483.9105 & .0712 & $-$13.7514 \\
03 & 3176.9600 & 4479.4869 & .0706 & $-$13.7721 & 3198.7869 & 4481.8681 & .0822 & $-$13.7836 \\
04 & 3176.9733 & 4479.4917 & .0642 & $-$13.7539 & 3198.7938 & 4481.8611 & .0747 & $-$13.7483 \\
05 & 3184.4585 & 4459.9941 & .0569 & $-$13.7644 & 3203.3598 & 4461.1151 & .0758 & $-$13.7225 \\
06 & 3184.4847 & 4459.9796 & .0662 & $-$13.7687 & 3203.3837 & 4461.0890 & .0822 & $-$13.5614 \\
07 & 3192.5618 & 4453.7872 & .0606 & $-$13.7936 & 3210.5695 & 4454.5485 & .0789 & $-$13.7416 \\
08 & 3192.5604 & 4453.7906 & .0657 & $-$13.7925 & 3210.5842 & 4454.5426 & .0714 & $-$13.7596 \\
09 & 3214.0726 & 4447.1540 & .0486 & $-$13.7717 & 3230.2991 & 4447.1838 & .0573 & $-$13.8035 \\
10 & 3214.0726 & 4447.1658 & .0520 & $-$13.7789 & 3230.3068 & 4447.2018 & .0745 & $-$13.7369 \\
11 & 3242.2729 & 4452.4620 & .0560 & $-$13.7351 & 3255.6101 & 4451.3406 & .0756 & $-$13.7226 \\
12 & 3242.3008 & 4452.4571 & .0636 & $-$13.7869 & 3255.6356 & 4451.3474 & .0668 & $-$13.7521 \\
13 & 3247.3892 & 4456.3960 & .0562 & $-$13.7746 & 3259.0011 & 4454.6361 & .0623 & $-$13.7179 \\
14 & 3247.3943 & 4456.4028 & .0615 & $-$13.7785 & 3259.0115 & 4454.6309 & .0709 & $-$13.7319 \\
15 & 3247.0500 & 4454.9440 & .0645 & $-$13.7750 & 3257.0941 & 4452.5905 & .0723 & $-$13.8075 \\
16 & 3247.0505 & 4454.9441 & .0696 & $-$13.7775 & 3257.0982 & 4452.5800 & .0804 & $-$13.7487 \\
17 & 3254.4639 & 4435.7605 & .0700 & $-$13.7896 & 3261.1670 & 4432.2108 & .0666 & $-$13.7592 \\
18 & 3254.4500 & 4435.7624 & .0735 & $-$13.7754 & 3261.1665 & 4432.2065 & .0804 & $-$13.7109 \\
19 & 3264.2620 & 4428.4694 & .0685 & $-$13.8066 & 3269.8034 & 4424.4836 & .0635 & $-$13.8043 \\
20 & 3264.2910 & 4428.4512 & .0777 & $-$13.7803 & 3269.8294 & 4424.4669 & .0838 & $-$13.7612 \\
21 & 3283.9466 & 4422.6495 & .0872 & $-$13.8059 & 3287.6769 & 4418.0644 & .0925 & $-$13.7280 \\
22 & 3283.9666 & 4422.6405 & .0800 & $-$13.8129 & 3287.6971 & 4418.0447 & .0858 & $-$13.7983 \\
23 & 3317.7075 & 4430.3043 & .0729 & $-$13.8042 & 3314.7639 & 4423.5710 & .0730 & $-$13.7451 \\
24 & 3317.7254 & 4430.3247 & .0736 & $-$13.7757 & 3314.7844 & 4423.5749 & .0851 & $-$13.7410 \\
& & & & & & & & \\                                                                                  
25 & 3177.4511 & 4480.9871 & .0438 & $-$12.5274 & 3200.5524 & 4483.9125 & .0326 & $-$11.8824 \\
26 & 3176.9679 & 4479.5222 & .0549 & $-$12.4756 & 3198.7734 & 4481.8883 & .0542 & $-$11.8233 \\
27 & 3184.4647 & 4459.9898 & .0608 & $-$12.5185 & 3203.3521 & 4461.1134 & .0353 & $-$11.8406 \\
28 & 3192.5469 & 4453.8118 & .0426 & $-$12.5346 & 3210.5591 & 4454.5391 & .0418 & $-$11.8384 \\
29 & 3214.0780 & 4447.1755 & .0514 & $-$12.5449 & 3230.3074 & 4447.1921 & .0474 & $-$11.8699 \\
30 & 3242.3070 & 4452.4713 & .0514 & $-$12.4980 & 3255.6328 & 4451.3491 & .0495 & $-$11.8312 \\
31 & 3247.4004 & 4456.4198 & .0377 & $-$12.5136 & 3259.0075 & 4454.6286 & .0336 & $-$11.7717 \\
32 & 3247.0462 & 4454.9560 & .0514 & $-$12.5031 & 3257.0802 & 4452.6036 & .0472 & $-$11.8295 \\
33 & 3254.4352 & 4435.8009 & .0692 & $-$12.5472 & 3261.1611 & 4432.2221 & .0551 & $-$11.8479 \\
34 & 3264.2919 & 4428.4862 & .0422 & $-$12.5278 & 3269.8101 & 4424.4895 & .0492 & $-$11.8609 \\
35 & 3283.9286 & 4422.6551 & .0405 & $-$12.5847 & 3287.6626 & 4418.0566 & .0542 & $-$11.8742 \\
36 & 3317.7059 & 4430.3364 & .0445 & $-$12.4950 & 3314.7809 & 4423.6060 & .0438 & $-$11.8031 \\
& & & & & & & & \\
\hline
\end{tabular}
\label{tabXYdtr}
\end{table*} 
%

It  is not  clear how  and why  the accuracy  of the  simultaneous fit
solution is  degraded using a partial  arc of the orbit.  We suspect a
complicated   interplay  with   CTE  residuals,   field-targets  color
dependencies in  the PSFs, or  just the deviations from  the absolutes
positions, all of which might propagate in the solution.

Therefore, until distances with  \textit{Gaia} for the reference field
stars or data covering more of the orbital phase will be available, we
now abandon the simultaneous fit of (absolute) astrometric and orbital
parameters, and explore in the next  section a simpler and more robust
(relative) approach.

\section{METHOD\,B: Two-Step Determination of Astrometric and Orbital Parameters}

In a two-body system, at any time, the baricenter (indicated hereafter
with G)  lies along the segment  connecting the two components  of the
system, and the position of G along  this segment is fixed by just one
parameter, the mass ratio (hereafter, $q$).

In this second  approach we obtain our solution in  two steps.  In the
first  step we  determine  $q$  and the  astrometric  parameters of  G
(positions, proper  motions and parallax),  and in the second  step we
solve  for the  relative  orbit, now  using just  the  Luh16\,A and  B
relative positions. Note that  the derived astrometric parameters will
be  in  the  relative  astrometric   system  of  the  reference  frame
$(2X,2Y)$, which is not an absolute system.
The orbital solutions  will not be affected by  CTE residuals, because
when looking at  differences between A and B positions  the effects of
CTE should cancel-out at a great level of accuracy.
The same is true for the  PSFs' color-dependencies, as Luh\,16 A and B
have essentially the same color and they would be affected by the same
systematic errors.

\subsection{Step\,1: Determination of Positions, Parallax, Proper Motions, and Mass-Ratio}
\label{s:step1}

In Fig.\,\ref{stack}  we can see in  one shot all the  projected space
motions during the  first 12 epochs of our  \textit{HST\,} campaign in
the  period between  August  22  2014 and  October  4  2016, for  both
Luhman\,16 A and B.
Not   all  of   the  components   in  all   the  epochs   are  clearly
distinguishable in this stacked image.
More clear is Fig.\,\ref{cycloid} where we show the complete series of
observed  positions   on  the   same  reference  frame   $(2X,2Y)$  of
Fig.\,\ref{stack}, indicating  Luhman\,16\,A with a filled  dot, and B
with an open circle. To better identify the epoch-pair, we connect the
AB components  with a black  segment.  Note  that at each  given epoch
(i.e., one single-orbit \textit{HST\,} visit) there are actually three
dithered individual observations taken less  than an hour apart, which
we can safely consider collected at the same astrometric epoch.

From the  observed 36 2D-data points  we would like to  derive for the
baricenter  of  the  Luhman\,16\,AB   system  ($G$)  five  astrometric
parameters:     its     positions     $(X_G,Y_G)$,     its     motions
$(\mu_{X_G},\mu_{Y_G})$,  and  most  importantly the  system  parallax
$\pi$.
However, the  baricenter $G$  is not  an observable,  but needs  to be
inferred from the relative positions of A and B components along their
mutual orbit.  So,  we need to add to the  unknown parameters the mass
ratio $q$ defined as  $\mathcal{M}_{\rm B}/\mathcal{M}_{\rm A}$ (i.e.,
$0<q<1$, where $\mathcal{M}$ indicate the mass of each component).
In the following we will describe  the procedure followed to fit these
six parameters.

By  virtue   of  the   principle  that   any  transformation   of  the
observational data  degrades them, while  numerical models do  not, we
perform this  numerical fitting process directly  in the observational
plane $(2X,2Y)$.
To accomplish our fit, we proceed iteratively solving for the best fit
of the data,  then de-trending for CTE residuals,  and finally fitting
again the CTE-de-trended data.

To  predict  the  position  of  the baricenter  we  make  use  of  the
sophisticated  tool  by  U.S.\ Naval  Observatory,  the  \textit{Naval
  Observatory Vector Astrometry Software}, hereafter NOVAS\footnote{
\texttt{http://aa.usno.navy.mil/software/novas/novas\_f/\-novasf\_intro.php}
}
(in version F3.1, Kaplan et al.\ 2011), which accounts for many subtle
effects,  such as  the accurate  Earth orbit,  perturbations of  major
bodies, nutation of the Moon-Earth system, etc.
We  are not  interested in  the absolute  astrometric calculations  of
NOVAS but only in the relative  effects. In computing the positions we
used an  auxiliary star  with no  motion and  zero parallax  (i.e., at
infinite distance), and finally compute the difference with respect to
our targets.
We  then  use a  Levenberg-Marquardt  algorithm  (the FORTRAN  version
\texttt{lmdif} available under MINIPACK, Mor\'e et al.\, 1980) to find
the minimization of six parameters: $X_G,Y_G,\mu_{X_G},\mu_{Y_G},\pi$,
and $q$.

%
\begin{table}
\caption{Astrometric parameters and mass-ratio of Luhman\,16\,AB. }
\center
\begin{tabular}{lcc}
\hline
$q$ = $\mathcal{M}_{\rm B}/\mathcal{M}_{\rm A}$                       &  0.848             & $\pm$ 0.023 \\
$\alpha_{\rm J2000.0}$ [hours]   & 10.82187776 & $\pm$ 65\,mas \\
$\delta_{\rm J2000.0}$ [degrees] & $-$53.3193958 & $\pm$ 45\,mas \\
$\mu_{\alpha\cos{\delta}_{\rm J2000.0}}$ [mas yr$^{-1}$]   & $-$2762.2 & $\pm$ 2.3 \\
$\mu_{\delta_{\rm J2000.0}}$ [mas yr$^{-1}$]  &  354.5   & $\pm$ 2.8 \\
$\pi$ [mas]  & 501.118     & $\pm$ 0.093 \\
$\varpi=\pi+0.28\pm0.01$  [mas]  & 501.398     & $\pm$ 0.093 $\pm$0.01 \\
\hline
\end{tabular}
\label{tabASTR}
\end{table} 
%


The first  iteration produced  a solution with  residuals as  large as
2\,mas, which were significantly larger  than our expected errors, and
most importantly clearly and simply  correlated with the $x^{\rm raw}$
and $y^{\rm  raw}$ coordinates. We  ascribed these residual  mainly to
CTE, but potentially also to differential chromatic and PSFs residuals
correlated with roll-angles.   The best fit is shown with  a blue line
in Fig.\,\ref{cycloid}.
A simple fit to the  residuals of the Observed$-$Calculated (O$-$C) of
baricenter  $2X$ and  $2Y$ as  function of  $x^{\rm raw}$  and $y^{\rm
  raw}$  deliver a  correction  that  can be  applied  to the  $(X,Y)$
positions. We  refer to these  de-trended positions  for both A  and B
components  with the  symbols $(X^{\rm  dtr},Y^{\rm dtr})$,  which are
given in Table\,\ref{tabXYdtr}.
After this  correction the (O$-$C) residuals  are perfectly consistent
with those expected for stars  of this luminosity for WFC3/UVIS, i.e.,
0.008 pixels, or, 320\,$\mu$mas.   Therefore we impute these remaining
residuals to just random errors.

\begin{figure}
\begin{center}
\includegraphics[width=89mm]{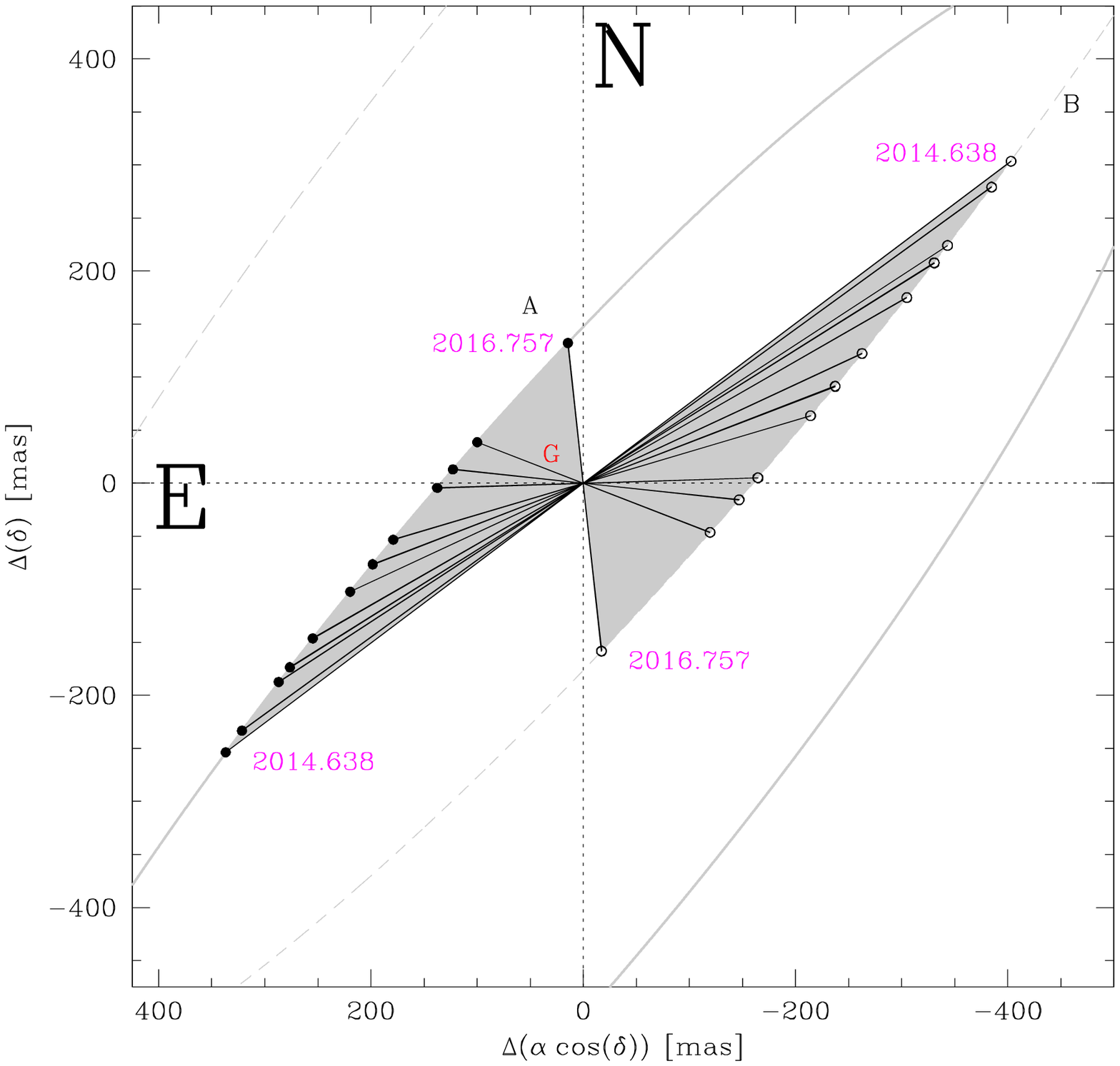}
\caption{
  In equatorial  coordinates, the positions  of A and B  components of
  Luhman\,16  relative  to our  estimate  of  their common  baricenter
  (indicated  with G).  We  have also  indicated the  arc  of the  two
  baricentric orbits  (solid line for  A, and  dashed line for  B) and
  shaded   the  orbital   area  mapped   during  this   \textit{HST\,}
  campaign. Note the over 120$^\circ$ covered by the projection of the
  true anomaly.
\label{absORB}
}
\end{center}
\end{figure}

\begin{figure*}
\begin{center}
\includegraphics[width=178mm]{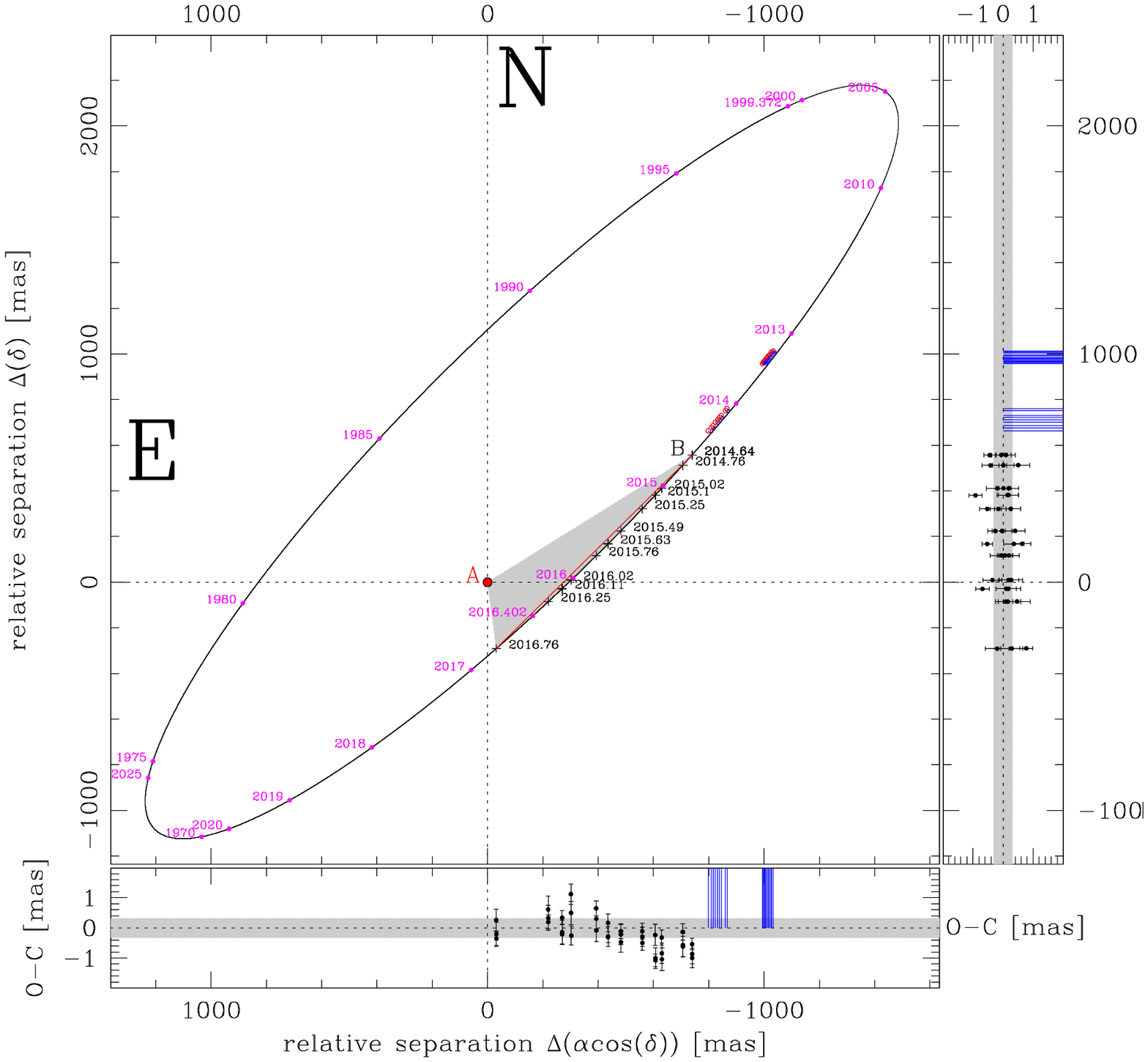}
\caption{
  The relative  orbit of Luhman\,16\,B  around A. To  better highlight
  the amount of curvature observed with \textit{HST\,} ($+$symbols) in
  the A-B relative orbit, a red  line connects the last with the first
  of the observed  relative positions for the  B-component.  The lower
  and right panels  show in mas the (O$-$C) residuals  with respect to
  our best  orbital model. A gray  band is given for  reference as the
  expected  uncertainty  for  any  individual  positional  measurement
  (0.32\,mas).   Red circles  show  the positions  available from  the
  Sahlman \&  Lazorenko 2015, they are  all out of scale  in the O$-$C
  panels.
  Blue  lines connect  the individual  SL15 data  points with  their
  expected  positions  on our  orbital  solution  at the  epoch  those
  observations were collected.
  See  Fig.\,\ref{relORBzoom} for  a more  meaningful comparison  of
  residuals.
\label{relORB}
}
\end{center}
\end{figure*}
%

%
Our final  astrometric solution is given  in Table\,\ref{tabASTR}.  To
assess  uncertainties we  conservatively used  the method  of residual
permutation, and take the 68-th  percentile of the deviations from the
mean as the 1-$\sigma$ error.
The relative parallax  (which we indicate with $\pi$)  is with respect
to the most  distant objects in the field (i.e.,  those that moved the
least)  and therefore  $\pi$ is  only a  lower limit  to the  absolute
parallax $\varpi$.
In  a   first  approximation  we  can   use  the  relative-to-absolute
correction, which was derived by SL15 as 0.28$\pm$0.01\,mas.
In Table\,\ref{tabASTR}  we also give  the value derived  for $\varpi$
employing this correction.
However,     we      note     that     our      relative     parallax,
$\pi=501.118\pm0.093$\,mas,  is already  \textit{significantly} larger
than     the    absolute     parallax     by     SL15,    which     is
$\varpi=500.51\pm0.11$\,mas, therefore implying  a closer distance for
Luh\,16\,AB, at no more than 1.9955$\pm$0.0004\,pc.
When  \textit{Gaia}   parallaxes  for  reference  stars   will  become
available    we   will    be   able    to   determine    an   accurate
relative-to-absolute correction for our $\pi$ value.

%
\subsection{Step\,2: Determination of the Orbital Parameters}
%
\label{sect:OP}

With $q$ and  the astrometric parameters of G derived  in the previous
step, the  position of  the baricenter  is known  at any  given epoch.
This enables us to plot in Fig.\,\ref{absORB} the observed data points
in equatorial  coordinates relative to the  baricenter, revealing that
our data  cover over 120$^\circ$  of the projected true  anomaly along
the orbits (i.e., $\sim$1/3 of the projected orbit).
%
We indicate the  area swept during our observations  with gray regions
for  both A  and B  components. For  reference, we  also show  the A-B
mass-reduced orbits for the solution derived later in this section.
The relative  positions, in  equatorial coordinates,  of Luhman\,16\,B
with  respect to  Luhman\,16\,A  are shown  in Fig.\,\ref{relORB}  and
given in Table\,\ref{XYpositions}.
%
\begin{table}
\caption{Cartesian relative equaorial positions of B with respect to A with estimated errors, in mas.}
\center
\begin{tabular}{lcccc}
\hline
\#ID & Julian Year & $\Delta\alpha\cos{\delta}$ & $\Delta\delta$ & $\sigma_{\Delta}$\\
     &             & [mas]                      & [mas]          &    [mas]       \\                                            
\hline
01 & 2014.63801 & $-$740.243 &   557.536 & 0.319 \\
02 & 2014.63820 & $-$740.047 &   557.310 & 0.319 \\
03 & 2014.75865 & $-$706.509 &   513.032 & 0.383 \\
04 & 2014.75873 & $-$706.521 &   512.498 & 0.348 \\
05 & 2015.02328 & $-$630.386 &   411.781 & 0.335 \\
06 & 2015.02347 & $-$630.535 &   411.335 & 0.373 \\
07 & 2015.10044 & $-$606.637 &   381.725 & 0.352 \\
08 & 2015.10053 & $-$607.376 &   381.722 & 0.343 \\
09 & 2015.25312 & $-$559.585 &   321.314 & 0.266 \\
10 & 2015.25321 & $-$559.728 &   321.680 & 0.321 \\
11 & 2015.49489 & $-$482.563 &   224.543 & 0.333 \\
12 & 2015.49507 & $-$482.249 &   224.899 & 0.326 \\
13 & 2015.63458 & $-$435.605 &   168.446 & 0.297 \\
14 & 2015.63467 & $-$436.025 &   168.136 & 0.332 \\
15 & 2015.75923 & $-$393.198 &   117.008 & 0.342 \\
16 & 2015.75932 & $-$393.531 &   116.713 & 0.376 \\
17 & 2016.01960 & $-$301.470 &     9.767 & 0.342 \\
18 & 2016.01969 & $-$302.055 &     9.818 & 0.385 \\
19 & 2016.11147 & $-$269.974 & $-$28.218 & 0.330 \\
20 & 2016.11167 & $-$269.841 & $-$28.225 & 0.404 \\
21 & 2016.24932 & $-$219.280 & $-$84.655 & 0.449 \\
22 & 2016.24948 & $-$219.498 & $-$85.021 & 0.415 \\
23 & 2016.75726 &  $-$31.285 &$-$290.547 & 0.365 \\
24 & 2016.75735 &  $-$31.701 &$-$291.065 & 0.398 \\
   &            &            &           &       \\                                            
25 & 2014.63805 & $-$739.763 &   556.973 & 0.193 \\
26 & 2014.75869 & $-$706.069 &   512.088 & 0.273 \\
27 & 2015.02332 & $-$629.855 &   411.596 & 0.249 \\
28 & 2015.10048 & $-$607.463 &   380.640 & 0.211 \\
29 & 2015.25316 & $-$559.945 &   320.916 & 0.247 \\
30 & 2015.49494 & $-$482.185 &   224.290 & 0.252 \\
31 & 2015.63462 & $-$436.057 &   167.271 & 0.179 \\
32 & 2015.75927 & $-$392.827 &   116.847 & 0.247 \\
33 & 2016.01964 & $-$302.832 &     9.213 & 0.313 \\
34 & 2016.11159 & $-$269.388 & $-$29.052 & 0.229 \\
35 & 2016.24936 & $-$219.672 & $-$85.045 & 0.239 \\
36 & 2016.75731 &  $-$31.870 &$-$290.080 & 0.221 \\
\hline
\end{tabular}
\label{XYpositions}
\end{table} 
%

~\\ 

To determine the  seven orbital parameters of visual  binaries we will
follow the trial-and-error approach described by Pourbaix (1994).
We employed the two software tools  developed and maintained by one of
us  (D.P.):  \texttt{union}  and  \texttt{epilogue}  (Pourbaix  1998).
Practical details about the  internal algorithms and some applications
are available in Pourbaix (1998a,b, and 2000).
Briefly,  \texttt{union} and  \texttt{epilogue} are  used together  to
simultaneously    adjust    the     observations    of    double-lined
spectroscopic-visual binaries (VB-SB2).
\texttt{union}  undertakes a  global optimization  then followed  by a
preliminary local search.  In order  to increase the chance of getting
the right  minimum, the user supplies  with a time range  to which the
orbital period is assumed to belong.
\texttt{epilogue}   processes  the   results  of   \texttt{union}  and
generates some  statistical information as  well as ephemerid  for the
solution.

%
\begin{table}
\caption{Derived orbital parameters and masses of Luh\,16\,AB.}
\center
\begin{tabular}{lccc}
\hline
parameter & min value & max value & weighted-mean $\pm$$\sigma$ \\
\hline
$a$ [arcsec] &   1.69  &  2.84 &  1.91$\pm$0.25   \\
$a$ [AU]     &   3.71  &  5.67 &  3.81$\pm$0.50   \\
$i$             [deg]    &  78.51  & 80.10 & 79.21$\pm$0.45   \\
$\omega$        [deg]    &  84     & 168$^*$(307)   & 107$^*$$\pm$18      \\
$\Omega$      [deg]    &  127.6  & 132.2$^*$(311) & 130.3$^*$$\pm$1.1    \\
$e$                      &  0.31   &  0.61 & 0.46$\pm$0.06  \\
$P$             [yr]     &  24.5   &  64.4 & 31.3$\pm$7.9    \\
$T_{\circ}$ [Julian\,yr]   & 2016.2  & 2018.4 & 2017.1$\pm$0.7   \\
$\mathcal{M}_{\rm tot}^\dagger$ [$\mathcal{M}_{\odot}$] & 0.044 & 0.085 & 0.056$\pm$0.020\\ 
$\mathcal{M}_{\rm tot}^\dagger$ [$\mathcal{M}_{\rm J}$] & 46 & 89 & 59$\pm$21\\ 
$\mathcal{M}_{\rm Luh\,A}^\ddagger$ [$\mathcal{M}_{\rm J}$] & 26 & 48 & 32$\pm$11\\ 
$\mathcal{M}_{\rm Luh\,B}^\ddagger$ [$\mathcal{M}_{\rm J}$] & 21 & 41 & 27$\pm$10\\ 
\hline
\end{tabular}
$^*$ 5-$\sigma$ clipped, rejecting 4 outliers. \\
Total mass, $^\dagger$ $\mathcal{M}_{\rm tot} = \mathcal{M}_{\rm Luh\,16\,A} +
\mathcal{M}_{\rm Luh\,16\,B} = a^3/P^2$.\\
$^\ddagger$ Assuming $q=0.848$, $\mathcal{M}_{\rm Luh\,16\,A} = 1/(1+q) \mathcal{M}_{\rm tot}$, and 
$\mathcal{M}_{\rm Luh\,16\,B} = q/(1+q) \mathcal{M}_{\rm tot}$,
\label{tabORB}
\end{table} 
%

The resulting orbital solution for the Luhman\,16\,A-B system is given
in  Table\,\ref{tabORB}.   To  estimate   uncertainties  we   use  the
conservative  approach of  permuting  the 36  residuals.  We give  the
minimum and  maximum values of  the different realizations,  and their
weighted averages.  We indicate in  the case of $\omega$  and $\Omega$
that a 5-$\sigma$ clipping exclude 4 outliers.

\begin{figure*}
\begin{center}
\includegraphics[width=178mm]{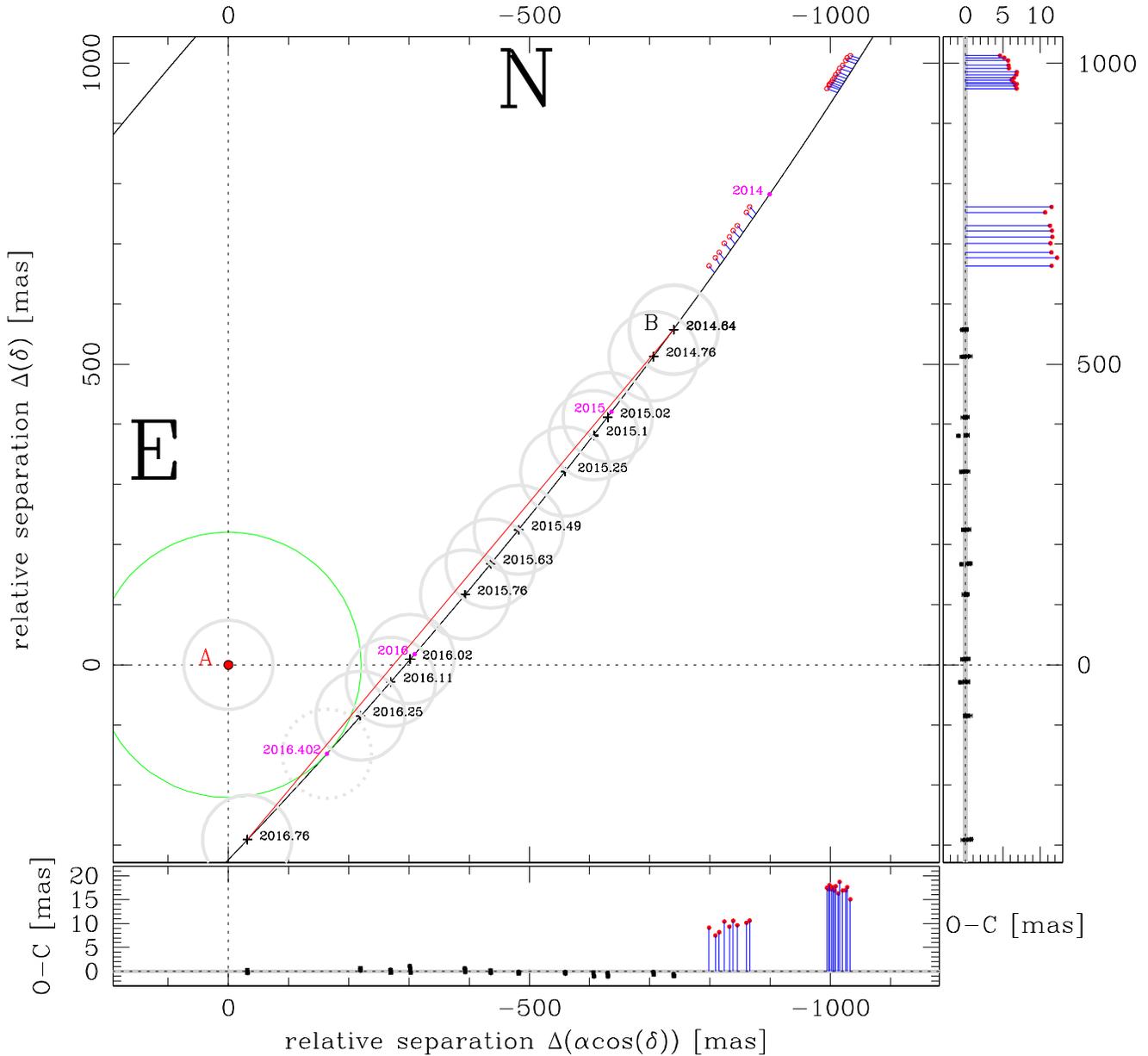}
\caption{
  Same as in Fig.\,\ref{relORB}, only focused on the observed data points
  from this work and from SL15.  The residuals are now shown in a
  scale that enable to see all residuals of the SL15 data points.  Note
  that neither a simple rotation or shifts, nor a change in the absolute scale
  can bring into agreement SL15's \textit{VLT\,} data points with our
  \textit{HST\,}'s data-set.  A green circle indicates the apparent
  periastron (at 220.5\,mas) estimated to have happened in
  2016.402. For reference, the size of the nominal FWHM with WFC3/UVIS
  filter F814W (74\,mas) is indicated with a gray thick circle at
  positions where A and B components were observed, and dotted at the
  not-observed apparent periastron.
\label{relORBzoom}
}
\end{center}
\end{figure*}

\subsection{Disagreement with SL15 astrometry}
\label{SL15dis}

We  have downloaded  the on-line  supporting information  available at
MNRAS,   for   Table\,A1  of   the   Sahlmann   \&  Lazorenko   (2015)
study.\footnote{
\texttt{http://mnrasl.oxfordjournals.org/content/453/1/L103/\-suppl/DC1}
}

This table provides the equatorial  coordinates of Luhman\,16\,A and B
collected with FORS2@\textit{VLT}  in 22 epochs between  April 14 2013
and  May  18 2014.  The  positions  are given  in  the  ICRF, after  a
correction  for  differential  color   refraction  (DCR)  effects  was
applied.

These  data-points   could  potentially  extend  by   1.36  years  our
\textit{HST}  data time  base-line, resulting  in better  estimates of
both astrometric  and orbital parameters. Extra  monitoring would also
improve the search for perturbations induced by third bodies.

We show the  relative positions transformed into  the tangential plane
in Fig.\,\ref{relORB} and Fig.\,\ref{relORBzoom} as small red circles.
Apparently the  SL15 data points  are not consistent with  our orbital
solution.
We attempt to  re-derive the orbital fit including also  SL15 data but
with no success.
No simple rotation nor shift can solve the inconsistency.  However, if
we assume the pixel  scale in SL15 to be off  by $\sim$5$\%$, the data
would be  nearly aligned  with our orbital  solution.  But,  even when
this is  assumed, and the  orbit is re-derived using  our \textit{HST}
data points and simultaneously including SL15 data points, the results
are not physically consistent.
In summary, we are unable  to fit simultaneously our \textit{HST} data
and those by SL15 collected at FORS2@\textit{VLT}.

\begin{figure}
\begin{center}
\includegraphics[width=88mm]{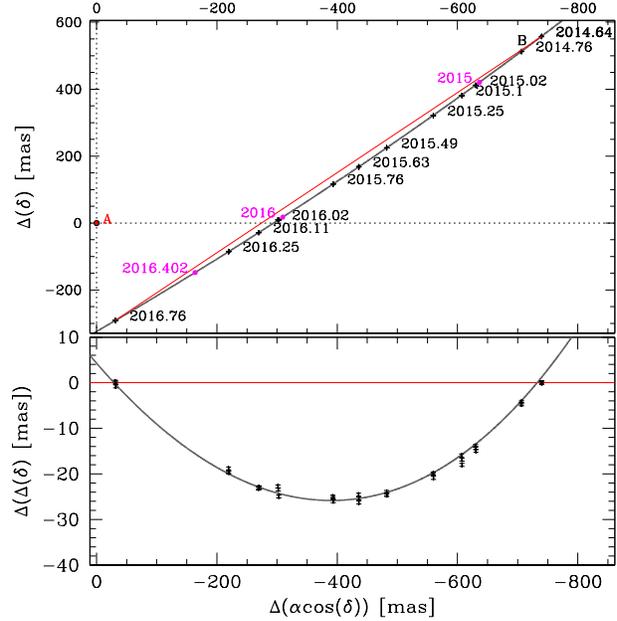}
\caption{
  Same as in Fig.\,\ref{relORBzoom}, only focused on the curvature. 
\label{curvature}
}
\end{center}
\end{figure}

In  Figure\,\ref{relORBzoom}  it can  be  appreciated  that while  our
observed  \textit{HST}  data  points significantly  deviate  from  the
straight line (by over 25\,mas, at $\sim$100\,$\sigma$) those observed
by SL15 are almost consistent with  a straight line, deviating no more
than very few mas, which seems  consistent with the observed amount of
scatter in their 22 data points. This is also consistent with the fact
that while our \textit{HST} data are exploring over 120$^\circ$ of the
projected true anomaly $\nu$, the SL15  data map only a few degrees of
$\nu$ (see Fig.\,\ref{relORB}).
This is  even clearer  in Fig.\,\ref{curvature},  where the  amount of
curvature traced by our \textit{HST} data is better exposed.

We  cannot  point  to  the exact  reason(s)  of  these  discrepancies,
however, we  suspect the usual limitations  of ground-based facilities
may be responsible, i.e.:
\textit{(i) } the limited parallax  factor covered by the ground-based
data (see Fig.\,3 of SL15);
\textit{(ii)} the FORS2 geometric  astrometric stability (camera focal
enhancing,  de-rotator, active  optic  stability) and  the effects  of
Earth's  atmosphere   on  the  astrometry  (variable   PSFs,  variable
extinction and residual DCR);
\textit{(iii)} the  limited angular  resolution for this  tight binary
($\sim$1\,arcsec   during   2013-2014,   i.e.,   comparable   to   the
ground-based seeing); and
\textit{(iv)} the  limited arc of  the AB orbit the  ground-based data
cover.
SL15 quote positional  errors within individual images as  low as 0.25
mas,  and  we  suspect  that  this  quoted  uncertainty  ---below  the
\textit{HST}  geometric  distortion  limit--- reflects  only  internal
errors,  which might  severely  underestimate the  true (internal  and
systematic)  errors. Nevertheless,  their  orbital solution,  although
poorly constrained by the data, still gives (O-C) residuals consistent
with their quoted errors.
It is also unclear whether in  deriving their proper motions they made
use of the  original WISE AB-photo-center position  (of the unresolved
components).  Instead, we  are only  using our  2014-2016 \textit{HST}
data,  as using  unresolved position  of  the photo-center  (and at  a
different wavelength)  would surely result in  decreasing the accuracy
of our solution, biasing our estimates for astrometric parameters of G
and $q$.

\begin{figure*}
\begin{center}
\includegraphics[width=88mm]{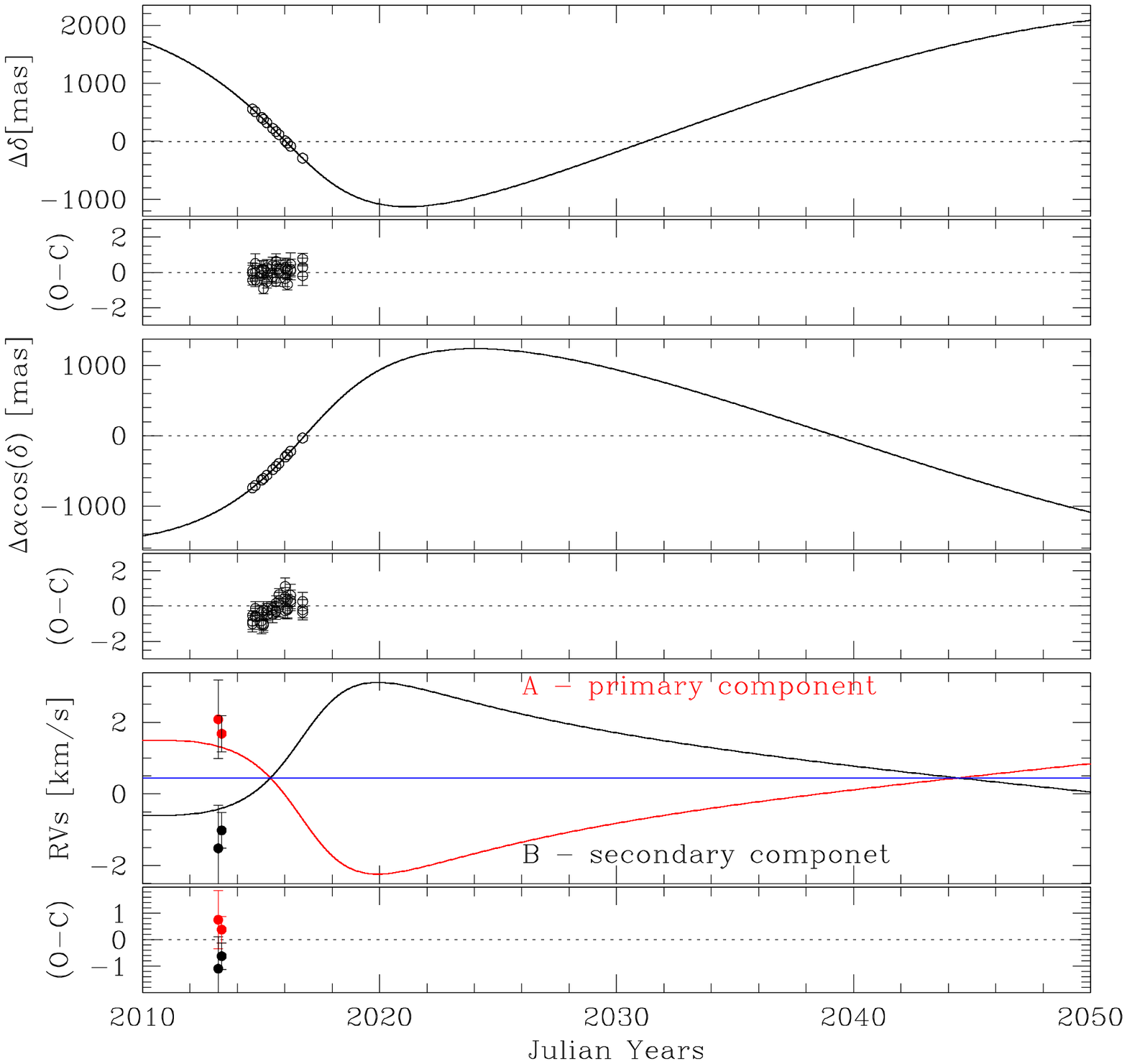} 
\includegraphics[width=88mm]{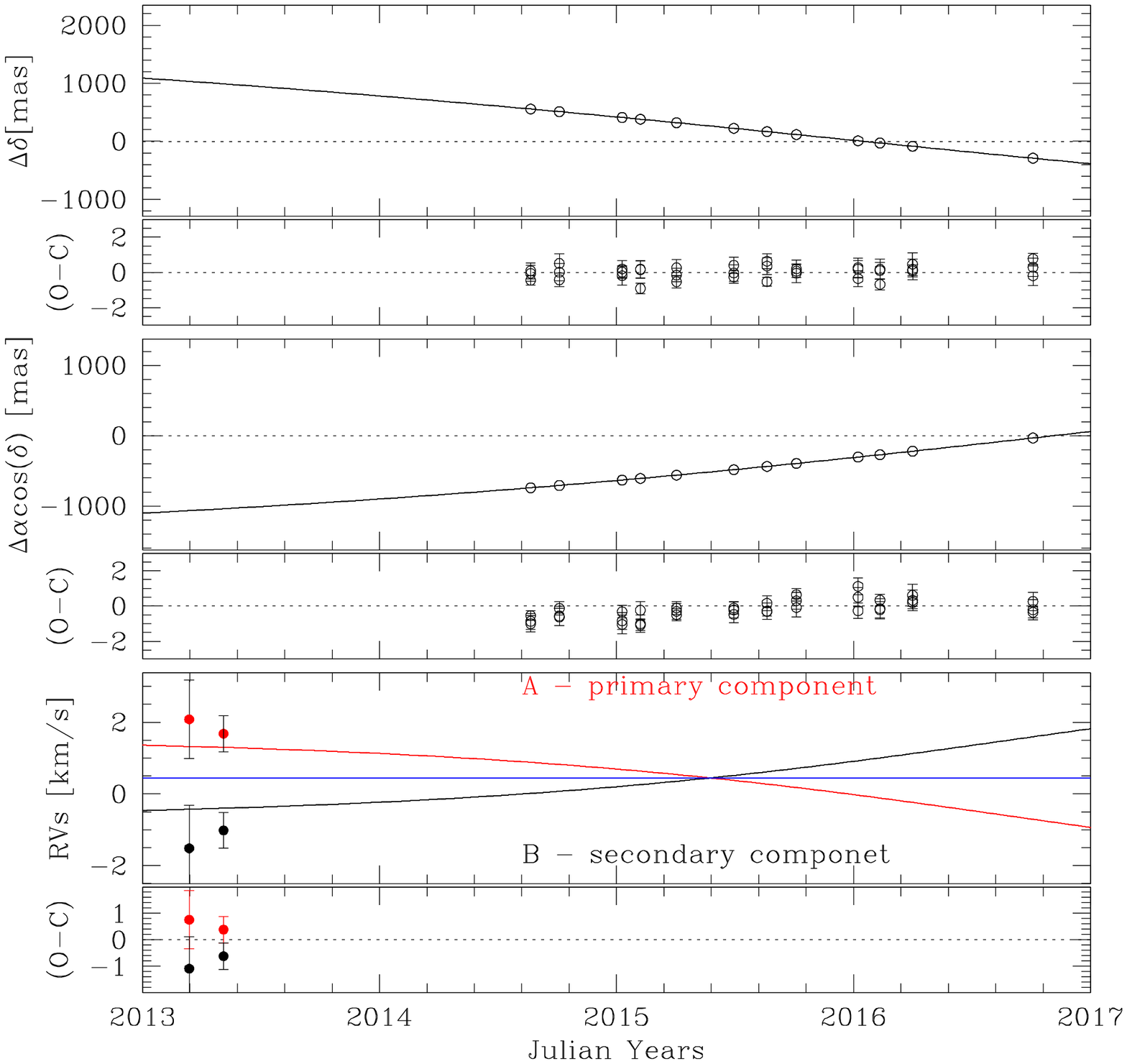}
\caption{
  From top  to bottom, the  three panels show the  comparisons between
  the observed  $\Delta(\delta)$, $\Delta(\alpha\cos{\delta})$, radial
  velocities  (RVs), and  the corresponding  quantities from  our best
  orbital  fit for  the  Luhman\,16\,AB system.  Each comparison  also
  shows  the Observed  $-$ Calculated  values below  the corresponding
  panels.
  The RVs available from literature were processed as described in the
  text.
  \textit{(Left:)} We show  the models over a full  A-B orbital period
  as reference for  the expected full amplitudes  of these quantities,
  altough observations are currently limited to just $\sim3.5$\,yr.
  \textit{(Right:)} Same plot,  but focused just on  the time interval
  where data are actually available so far.
\label{XYV}
}
\end{center}
\end{figure*}

\section{Comparison with Radial Velocities from the Literature}

In deriving  the orbit of  a visual  binary from astrometry  only, the
inclination (which appears as $\cos{i}$) is degenerate in $i$ and $-i$
and  only  radial velocities  (RVs)  can  remove this  degeneracy  and
determine the sign.
Thankfully  a  couple  of  RVs measurements  are  available  from  the
literature and  \texttt{epilogue} and  \texttt{union} can  process RVs
(Pourbaix 1998).

The  only  suitable  resolved  heliocentric  RVs  available  from  the
literature are  the ones of Kniazev  et al.\ (2013) and  Crossfield et
al.\ (2014).
Kniazev et al.\ at  epoch 2013.196 give 23.1$\pm$1.1\,km\,s$^{-1}$ and
19.5$\pm$1.2\,km\,s$^{-1}$, for components A and B respectively.
Crossfield al.\ at epoch  2013.342 give for 20.1$\pm$0.5\,km\,s$^{-1}$
and 17.4$\pm$0.5\,km\,s$^{-1}$, for components A and B respectively.

Inevitably, there  are numerous difficulties to  estimate the absolute
RV zero points of  heterogeneous observations collected with different
instruments at  different telescopes in  different sites, and  this is
particularly  true when  comparing  RVs derived  from  optical and  IR
observations.
Therefore, as we are interested mainly  in the relative RVs of the two
components,   we    assumed   the   mass   ratio    $q$   derived   in
Sect.\,\ref{s:step1}, and  compute the RVs  of the baricenter  for the
two references. We  then impose the velocity of the  baricenter in the
two data  sets to  be equal.\  This was obtained  imposing a  shift of
$+$2.6\,km\,s$^{-1}$ to the Crossfield et al.\ data.

In  Figure\,\ref{XYV}  we  compare   our  orbital  solution  with  our
astrometric  data  points and  with  the  available radial  velocities
processed as just described above.
From top  to bottom, we  show our  residuals as function  of $\delta$,
$\alpha^*$, and RVs.
These two  RV pairs are sufficient  to suggest that $i$  should have a
negative value.
Consequently also the $\omega$ and $\Omega$ change by 180$^\circ$. For
clarity the new amended values are given in Table\,\ref{tab-i}.
%

%
\begin{table}
\caption{Final orbital parameters and masses of Luhman\,16\,AB. }
\center
\begin{tabular}{lcc}
\hline
$a$ [arcsec] &   1.91   & $\pm$ 0.25   \\
$a$ [AU]     &   3.81   & $\pm$ 0.50   \\
$i$             [deg]    & $-$79.21   & $\pm$  0.45      \\
$\omega$        [deg]    & 287      & $\pm$ 18      \\
$\Omega$        [deg]    & 310.3    & $\pm$  1.1      \\
$e$                      &   0.463  & $\pm$  0.064     \\
$P$             [yr]     &  31.3    & $\pm$  7.9        \\
$T_{\circ}$ [Julian yr]  & 2017.12   & $\pm$  0.65  \\
$\mathcal{M}_{\rm tot}$ [$\mathcal{M}_{\odot}$] & 0.056 & $\pm$0.020\\ 
$\mathcal{M}_{\rm tot}$ [$\mathcal{M}_{\rm J}$] & 59    & $\pm$21\\ 
$\mathcal{M}_{\rm Luh\,A}^\ddagger$ [$\mathcal{M}_{\rm J}$]  & 32 & $\pm$11\\ 
$\mathcal{M}_{\rm Luh\,B}^\ddagger$ [$\mathcal{M}_{\rm J}$]  & 27 & $\pm$10\\ 
\hline
\end{tabular}
\label{tab-i}
\end{table} 
%

%

%
\section{Photometry}
%

%
Our  \textit{HST} photometry  were  calibrated to  the Vega  magnitude
system  by  adding  the   filter  zeropoints  (ZP$_{m_{\rm  F814W}}  =
29.02\pm0.02$ and ZP$_{m_{\rm F606W}} = 32.24\pm0.02$, following procedures in Bedin
et al.\ 2005)\footnote{
and zero points at http://www.stsci.edu/hst/wfc3/phot\_zp\_lbn} to the instrumental magnitudes given in Table\,\ref{tabXYdtr}.
The absolute accuracy of the calibration is about 0.02 mag per filter.
Our  observations  are not  well  suited  for constraining  rotational
modulation  of these  sources, but  provide some  of the  most precise
optical photometry for the two brown dwarfs.

\begin{figure}
\begin{center}
\includegraphics[width=88mm]{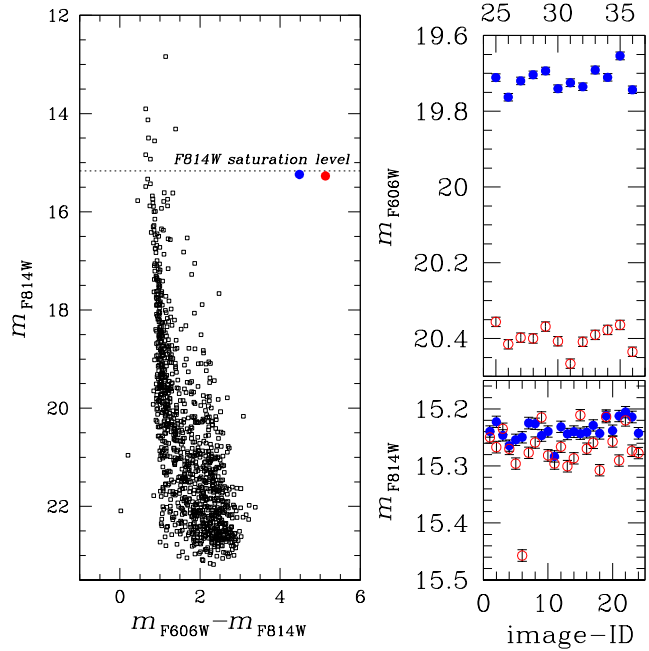} 
\caption{
  Summary  of the  photometric data  released  as part  of this  work.
  Black symbols are field objects observed within our WFC3/UVIS field,
  Luh\,16\,A is indicated in blue and  Luh\,16\,B in red.  On the left
  panel we  show the CMD of  all the sources.  A  dotted line indicate
  the on-set of saturation.
  On the right panels we show the individual magnitudes for Luh\,16\,A
  and B (F606W  on top, and F814W  on bottom), as measured  in each of
  the  36 images  used in  this work,  as function  of the  image \#ID
  (Table\,\ref{tabimg}).
\label{CMD}
}
\end{center}
\end{figure}

In  Fig.\,\ref{CMD}  we  summarize  the photometry  obtained  for  the
targets (Luh\,16\,A in blue, and B in  red) and for the sources in our
WFC3/UVIS field (black square symbols).
On the left panel we show the $(m_{\rm F606W}-m_{\rm F814W})\, vs.\,
m_{\rm F814W}$ color-magnitude diagram (CMD).
The onset of saturation in the 60\,s exposure in F814W is indicated by
a dotted line.
Note  how Luh\,16\,A  and B  are  considerably redder  than any  other
source in the field, and just below the saturation level (as planned).
On the right  panels we show our registered  and calibrated magnitudes
obtained for each image.
The error  bars associated with  each data  points is the  r.m.s.\ for
stars within 1 magnitude below the saturation level.  This is 0.01 for
all, with the exception of B in F606W, for which we associate an error
of  0.0125 mag.   Both Luh\,16  components  appear to  have a  scatter
considerably  larger than  field objects  of the  same magnitudes,  in
particular the B component.

Luhman\,16\,B  is one  of the  highest-amplitude variable  brown dwarf
known (Gillon et al.\ 2013, Biller et al.\ 2013, Buenzli et al.\ 2014,
2015) and  spatially resolved  studies discovered variability  also in
Luhman\,16\,A (Buenzli et al.\ 2015) emerging from the rotation of the
heterogeneous cloud deck of  the brown dwarf.  Time-resolved precision
photometry  by Gillon  et al.\  (2013) showed  rapidly evolving  light
curve  variations  in  Luhman\,16\,B  and  multi-band  photometry  and
resolved  ground-based  spectroscopy  by  Biller et  al.\  (2013)  and
Burgasser  et  al.\  (2013),  respectively,  showed  wavelength-  (and
therefore pressure-) dependent phase shifts and amplitude variations.
Buenzli  et  al.\  (2014,   2015)  used  high-precision  time-resolved
\textit{HST}  spectroscopy   (0.8-1.7  $\mu$m  range)  to   show  that
correlated temperature and cloud  thickness variations are responsible
for the  observed near-infrared  spectral evolution, similar  to other
variable L/T transition brown dwarfs (e.g., Radigan et al.\ 2012, Apai
et al.\ 2013).   Crossfield et al.\ (2014)  presented time-resolved CO
line profile observations to derive  the first Doppler-imaging map for
a  brown  dwarf, which  revealed  multiple  bright and  dark  patches;
Karalidi  et al.\  (2015) used  a Markov  Chain Monte  Carlo-optimized
forward modeling  procedure (\texttt{Aeolus}, with a  finely pixelized
sphere  and assuming  elliptical  structures) to  derive the  simplest
surface  brightness   map  that   is  consistent  with   the  observed
variations.  Although  taken  hundreds  of rotations  apart  and  with
different techniques, the Crossfield et al.\ and Karalidi et al.\ maps
show some similarity  in terms of the  approximate size, distribution,
and contrasts of the features.\\

The \textit{observed}  variability for component A  is 0.11 magnitudes
in  F606W and  0.08  magnitudes in  F814W, while  for  component B  we
observed  variations as  large as  0.11 magnitudes  in F606W  and 0.25
magnitudes  in   F814W.  These  are  significantly   larger  than  the
$\sim$0.01  photometric   precision  of   our  measurements.   As  our
observations  are not  resolved  on the  timescale  of the  rotational
periods of Luhman\,16\,A and B,  we cannot derive peak-to-peak changes
for the  objects, therefore the  reported values should  be considered
lower limits on the intrinsic amplitudes at these wavelengths.
Red-optical variability  has been  detected in  other T  dwarfs (e.g.,
Gillon et al.\  2013; Heinze et al.\ 2015), which  suggests that cloud
amplitudes are likely  as large if not larger in  the red optical than
they  are  in  the near-infrared  (1-1.6\,$\mu$m)  bands,  potentially
providing stronger  constraints on condensate grain  size distribution
(e.g., Lew et al.\ 2016).

Alternately, variability  of nonthermal  emission induced  by magnetic
activity could be responsible for this  signal, as such emission has a
much  greater  relative  impact   at  optical  wavelengths  where  the
photospheric  emission is  weak (e.g.,  Croll  et al.\  2016; Gizis  et
al.\  2017).  Magnetic  emission has  been detected  from T  dwarfs at
optical (e.g.,  Burgasser et al.\  2003; Kao  et al.\ 2016;  Pineda et
al.\  2016) and  radio  wavelengths (e.g.,  Route  \& Wolszczan  2012,
2016);  however,  neither component  of  Luhman\,16\,AB  was found  to
exhibit H$\alpha$ emission  to an equivalent width  limit of 1.5~{\AA}
(Faherty et al.\ 2014), and  the system was undetected in single-epoch
radio   and  X-ray   observations  (Osten   et  al.\   2015).  Further
investigation  into  the  variability   of  this  system,  over  broad
wavelength and time spans, is certainly warranted.

Our  observations  are also  consistent  with  the extended  base-line
(spatially  unresolved) photometric  monitoring of  the Luhman\,16\,AB
system in the Pan-STARRS-Z band and  in the SDSS $i'$ band, which show
that modulations 0.05 magnitude are typical over the timescales of the
rotational period, but  these authors also report changes  as large as
0.1\,mag (Street et al.\, 2015).


%

%
\section{Presence of Exoplanets}
%

The  main goal  of this  project  is to  search for,  and possibly  to
characterize, exoplanets and other bodies,  in P-type or S-type orbits
around the Luhman\,16\,A and B components.
We summarize the results of this preliminary work as follow:

(1)  We  inspected  each  epoch  for faint  companions,  and  find  no
detectable third body co-moving with the Luhman\,16\,A-B system within
our WFC3/UVIS  field of view. At  2\,pc, our search area  extends from
1~AU (0.$^{\prime\prime}$5) to 80~AU.
Assuming   a  detection   limit  of   $m_{\rm  F814W}$=23   magnitudes
(Fig.\,\ref{CMD}),  we can  rule  out well-resolved  companions to  an
absolute  magnitude  limit  of  M$_{\rm F814W}$  $\approx$  26.5.  For
comparison, the  T9 dwarf  UGPS~J072227.51$-$054031.2 has  an absolute
$i$-band  magnitude  of   26.2  (Leggett  et  al.\   2012),  so  these
observations rule out companions warmer  than a Y dwarf ($T_{\rm eff}$
$\gtrsim$\,500\,K).
It appears that  these new \textit{HST} images do not  provide any new
constraints (from direct  detection) that surpass those  from Melso et
al.\ (2015).\\

\begin{figure*}
\begin{center}
\includegraphics[width=150mm]{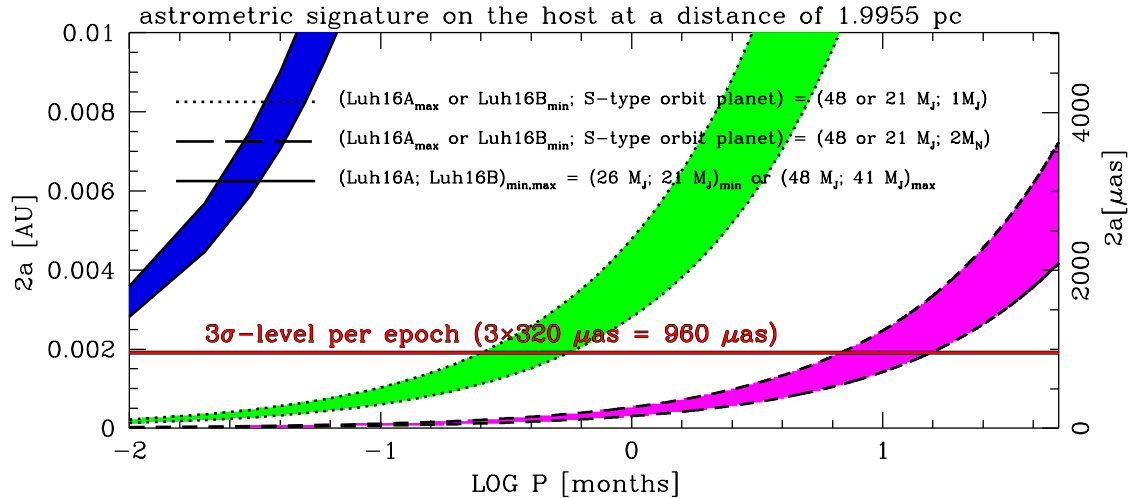}
\caption{
Assuming a distance of 1.9955\,pc  for Luhman\,16\,AB, and the maximum
and  mininum   masses  for   Luh\,16\,A  and  Luh\,16\,B   derived  in
Table\,\ref{tabORB}, respectively, we show  the relation between twice
the  semi-major axis  2a  of  the baricentric  orbit  of the  luminous
companions (A  or B)  around the  center of  mass with  a non-luminous
planet, and the period (P).
Dotted  lines  show  this  relation  for  a  planet  with  a  mass  of
1\,$\mathcal{M}_{\rm Jupiter}$.  The green  region indicates  the loci
for intermediate masses for A or B.
The dashed  lines (and corresponding  region in magenta)  indicate the
corresponding  astrometric signal  induced  by a  2\,$\mathcal{M}_{\rm
  Neptune}$ exoplanets.
For reference,  the solid lines (and  blue region) show the  amount of
A-B orbital motion if the system was on a circular orbit.
Twice the  semi-major axes  because that  is the  maximum size  of the
total range  of motion for a  circular orbit that we  can measure. The
horizontal red  line sets our  limit to the astrometric  signal, i.e.,
3\,$\sigma$$\simeq$960\,$\mu$as,  or, for  a distance  of $\sim$2  pc,
equivalent to $\sim$0.002\,AU (i.e., about 300\,000\,km).
\label{exop}
}
\end{center}
\end{figure*}

In Fig.\,\ref{exop} we show the  amplitude of the expected astrometric
perturbation  for  a number  of  cases,  over-imposed on  our  current
3\,$\sigma$ threshold.   For simplicity, we assume  circular planetary
orbits as  these are more likely  to be stable, and  adopt the maximum
and minimum masses for  Luh\,16\,A and Luh\,16\,B, respectively, given
in Table\,\ref{tabORB}.
 
(2) Using our  \textit{HST}-only data, we confirm the  results by SL15
as we also do  not see any significant residual in  our motions at the
$\sim$mas level  compatible with the  perturbation induced by  a third
body having  a mass  larger than  2 Jupiter  Masses ($\mathcal{M}_{\rm
  J}$), and period between 20\,days and 300\,days.

(3)  Our longer  time base-line  of  2.12\,yrs, compared  to the  SL15
base-line  of  1.09\,yrs,   allow  us  to  exclude   the  presence  of
2\,$\mathcal{M}_{\rm J}$ planets with period as long as 2\,yrs.

(4) We can also exclude  the presence of 1\,$\mathcal{M}_{\rm J}$ with
period between 20\,days and 2 years.

(5)  Finally, we  can extend  this claim  down to  significantly lower
masses, i.e.,  down to  Neptune masses for  periods longer  than about
1\,year.

To  increase sensitivity  and place  a lower  limit to  what we  could
detect, we can  average the residuals before and  after 2015.75, where
there  might seem  to be  a marginally  significant residual  of about
0.5$\pm$0.2\,mas (mainly in $\alpha^*$).
Even assuming that this can be  ascribed to a genuine signature caused
by  a third  body  ---rather than  a more  likely  effect of  residual
systematic  errors---  and assuming  masses  between  the maximum  and
minimum masses for Luh\,A and B, it would imply a period larger than 2
years and a mass not greater than one Neptunian mass.

(6) Neither  a simple shift,  nor a rotation,  nor a scale  change can
reconcile our \textit{HST}-derived positions  with those by SL15 based
on data obtained with \textit{VLT} from ground.  The two data sets can
\textit{not}  be  fit  with  a   common  orbital  solution  unless  an
\textit{ad-hoc} third  body is used  to explain that  discontinuity; a
possibility that at this stage we have not explored.

%
\section{Electronic Material}
%

As  part of  this  work, we  electronically  release as  supplementary
material on the Journal: \textit{i)} The individual observed positions
and  magnitudes  for  A  and   B  in  the  tables.   \textit{ii)}  The
astrometrized stack of the  \textit{HST} observations in filter F814W.
\textit{iii)} The photometry for the field objects.

%
\section{Conclusions}
%

In  this  paper  we  presented  the first  results  from  an  ongoing,
high-precision,  \textit{Hubble   Space  Telescope}-based  astrometric
monitoring campaign  targeting the Luhman\,16\,AB system,  the closest
brown dwarfs to the Sun.
The key findings of our study are as follows:

\textit{(i)} We rule out co-moving companions down to Y dwarf spectral
types in the range 1-80\,AU. This result confirms the finding of Melso
et  al.\   (2015),  which   is  based  on   superior  data   for  this
analysis. Note that Melso et  al.\ also extended this investigation at
wider separations.

\textit{(ii)} Our \textit{HST} data alone confirm the results by SL15:
we  find  no evidence  for  $\sim$mas-level  residuals that  would  be
consistent with  an exoplanet in the  system with a mass  greater than
2\,$\mathcal{M}_{\rm Jup}$ and with a period between 20 and 300 days.

\textit{(iii)}  We extend  this  results to  lower  masses and  larger
periods.   In  particular no  exoplanets  with  masses larger  than  a
Neptunian mass and period between 1 year and 2 years.

\textit{(iv)} Our measurements significantly  improve the parallax and
proper motion  of the Luhman\,16\,AB  system, and indicates  that this
system is closer to the Sun ($<$2 pc) than previous measurements.

\textit{(v)}  We have  also  improved on  the mass  ratio  $q$ of  the
Luhman\,16\,AB  system and  mapped  the motions  through the  apparent
periastron.

Further  significant  improvements  on the  astrometric  solution  and
parallax of the Luhman\,16\,AB system can be made in the near-future:
We plan to focus on  the trailed \textit{HST} images already collected
and those planned for August  2018, to further improve the astrometric
precision and search exoplanets down to few Earth masses.
Follow-up observations  with the  VLT/CRIRES+ instrument  will provide
accurate  radial velocity  data, an  important complement  to our  2-D
astrometric data,  as it will  provide the missed  component necessary
for  the complete  tri-dimensional  picture of  the  kinematic in  the
system.
In  addition, the  Gaia  DR2 dataset  will  provide absolute  motions,
positions, and distances of several stars in the field, allowing us to
link our positions to an absolute system.

\section{Acknowledgments}
We thank an anonymous referee for  the critical and careful reading of
our mauscript, and  for her/his suggestions and  corrections that have
contributed to improve this work.
D.A.  and A.B.  acknowledge  support from  STScI  grants GO-13748  and
GO-14330.   We warmly  thank  Shelly Meyett  and  Peter McCullough  at
STScI, our Program  Coordinator and Contact Scientist  for their great
support during the planning of the multi-year observations.

\noindent


%


\label{lastpage}


\end{document}